# Solvent-side observation on vibrational energy transfer by transient grating spectroscopy:

## Bridged azulene-anthracene

Hiroki Fujiwara[*][†]

Graduate School of Science, Kyoto University, Sakyo-ku, Kyoto 606-8502, Japan

Telephone number: (+81)75-753-4024

Fax number: (+81)75-751-4000

[*]Present addresses: Department of Chemistry, Durham University, South Road, Durham DH1 3LH, United Kingdom

[†]E-mail address: hiroki.fujiwara@durham.ac.uk, Phone number: +44 (0)191 334 2596

**ABSTRACT**

Transient grating acoustic spectroscopy has been applied to studies on the vibrational energy relaxation process of the electronic ground state of azulene, two 1-alkylazulenes, and five bridged azulene-anthracenes in three different solvents: 1,1,1-trichloro-1,2,2-trifluoroethane, acetonitrile, and xenon. The solute molecule was vibrationally excited by the photo-excitation of the auzlenyl group to the $S_1$ state through the fast internal conversion, and the rate of solvent thermalization due to the vibrational energy relaxation was determined. The thermalization rates for 1-alkylazulenes and bridged azulene-anthracenes were faster than that of azulene. Based on the results of the thermalization rates of 1-alkylazulenes, we concluded that the acceleration of the energy dissipation from the azulenyl group induced the faster energy dissipation from the solute to the solvent. The vibrational normal mode analysis suggests that the density of the vibrational modes and anharmonic coupling between the vibrational modes induce the faster intramolecular energy redistribution. In xenon, the solvent thermalization rates were close to the energy dissipation rates of the solute molecule reported using transient absorption spectroscopy. On the other hand, in the chlorofluorocarbon, the ratio of the thermalization rate to the dissipation rate was a dependent of the molecular species, and especially faster solvent thermalization rate of 9-(6-(azulen-1-yl)hexyl)anthracene was predicted. The vibrational characteristics of this solvent are discussed in the relation to the vibrational modes and structures of the bridged compounds.

**1. Introduction**

Vibrational Energy Relaxation (VER) is one of the earliest elemental steps of a chemical reaction process in solvent. The physical insight into VER dynamics is necessary for predicting and controlling chemical reactions in liquid phase.[1] However, some VER dynamics are not fully understood.[2]

In most spectroscopic studies on VER, dynamics of vibrationally excited solute molecule has been detected from the side of the photoexcited molecules, but not from the side of the solvent molecule which will accept the excess energy,[3] since it is generally quite difficult to detect the solvent molecules interacting with the solvent molecule among numerous bulk solvent molecules.[4] In order to examine the solute-solvent energy transfer, however, the solvent-side observation is more preferable, because of the two reasons. First, energy dissipation is more easily accessible by the solvent-side observation. Since the energy fraction of intermolecular energy transfer accompanying IVR process could be considerably smaller than that of CET process, the total energy flux should be detected. On the other hand, only the population of Frank-Condon active modes are detected by spectroscopic measurements on the solute molecule.[5] Second, the total kinetic information, such as the translational, rotational, and vibrational degrees of freedom, is more easily accessible using solvent-side observation as will be described in the following. Studies on the solute side as well as the solvent side are indispensable for understanding solvent commitment on IVR. There are several spectroscopic methods for solvent-side observation, such as IR transient absorption,[6] time-resolved Raman spectroscopy[7] and acoustic TG method.[8] The first two methods measure the population of solvent's vibrational modes, and mostly limited to the bulk solvent with vibrational modes specifically measurable due to the sensitivity. The third method (TG) is universal in terms of solvent species including supercritical fluids. TG method monitors the transient temperature of major energy accepting degrees of freedom, translation. Rotational degrees equilibrate with translation within the time resolution. Teramiza et al. have extensively studied the non-radiative process of the photo-excited molecules in solutions by applying the TG method.[9] Recently they measured translational temperature rise kinetics of supercritical fluids and organic solvents during VER of azulene by the TG method.[9g, 9h] They compared the time constants of the solvent temperature rise with the energy dissipation time of azulene determined by the transient absorption measurements of the identical system. They found that the time constant of the temperature rise was longer than the solute relaxation time constant, except xenon solution. They explained this delay as the VER time between solvent molecules after the V-V energy transfer from the solute to the solvent, and assessed the fraction energy flux carried by the V-V energy transfer in the total energy flux from the solute to the solvent.

An exceptional IVR phenomenon of bridged azulene-anthracenes was reported using the TA spectroscopy on the hot band.[10] They excited the azulenyl group to the electronic $S_1$ state using visible light (ca. 570nm) to generate vibrationally excited state localized to the azulenyl group through the fast internal conversion (< 1 ps). They observed local vibrational energies in azulenyl and anthracenyl groups separately to find three components with the identical kinetic coefficients in the two moieties. The fastest component (ca. 3 ps$^{-1}$) was assigned to the time propagation of vibrational normal modes. The second component (ca. 200 ns$^{-1}$) was identified as IVR, and the third component (ca. 30 ns$^{-1}$) was explained as the intermolecular energy transfer from solute to solvent molecules (CET). They showed that the kinetic coefficient of IVR components was independent of the solvent: 1,1,2-trichloro-1,2,2-trifluoroethane (CFC-113) or xenon. In this work, we studied the same systems in various solvents from the solvent side. Our aim is to explore the possibility of the energy dissipation from the solute to the solvent during the IVR process assigned by the TA study. By comparing the solvent thermalization rate with the solute CET rate, we can address the energy dissipation from the solute to the solvent.[41] Preliminary experimental results have been published previously for three kinds of bridged azulene-anthracenes and two kinds of 1-alkylazulenes in CFC-113 and acetonitrile.[11] Here we extended the system and the solvents; two kinds of 1-alkylazulenes, five kinds of bridged azulene-anthracenes dissolved in CFC-113, acetonitrile and xenon. We will discuss IVR issue with the help of molecular orbital calculations. The rest of paper is composed as follow. In Sec. 2, we describe the synthetic methods of bridged azulene-anthracenes, laser apparatus used for the measurements, and theoretical calculation details. Sec. 3 shows the experimental results. Typical acoustic TG signals are presented and the signal analysis will be presented. In Sec. 4, the thermalization kinetics and the intermolecular energy transfer during IVR process are examined. In Conclusions, our answer to the above question is shown.

**2. Experimental Methods**

  **A. Materials**

Solute compounds are azulene, 1-ethylazulene (AzEt), 1-dodecylazulene (AzDo), and five bridged azulene-anthracenes; 9-(2-(azulen-1-yl)ethyl)anthracene (AzEtAn), 9-(3-(azulen-1-yl)propyl)anthracene (AzPrAn), 9-(6-(azulen-1-yl)hexyl)anthracene (AzHeAn), 9-((azulen-1-ylmethoxy)methyl)anthracene (AzCOCAn) and 9-[2-(azulen-1-ylmethoxy)ethoxy)methyl]anthracene (AzCOCCOCAn) as shown in Scheme 1. Azulene was purchased (Nacalai Tesque) and purified twice by sublimation. AzCOCAn and AzCOCCOCAn were obtained from a collaborative laboratory and used without further purification. AzEt, AzDo, AzEtAn, AzPrAn and AzHeAn were synthesized and purified twice by column chromatography.[12] The details of the synthetic procedure are given in Supporting Information. Solvents were 1,1,1-trichloro-1,2,2-trifluoroethane (CFC-113, Fluka, >99.7%), acetonitrile (Nacalai Tesque, spectral grade) and xenon (Spectra Gases, 99.999%). All solvents were purchased and used without further purification.

### B. Transient grating acoustic method
#### (1) Theoretical Background

TG method is one of time-resolved four-wave mixing spectroscopic methods,[13] and for the present purpose we measure the acoustic signale.[9d-f] In essence, two pump pulses are incident on the sample solution simultaneously with a certain angle to generate interference fringe. Solute molecules in the solution are excited along this spatial pattern of light intensity. Due to the energy dissipation from the photo-excited molecule, the temperature of the solvent molecule will increase along the sinusoidal pattern. The temperature rise of the solvent induces the acoustic standing wave of the solution by the local heat expansion. This acoustic wave signal is generally called impulsive stimulated thermal scattering (ISTS). The time profile of this acoustic wave (acoustic TG) will be monitored by the intensity of the diffraction of the probe pulse which is introduced into the sample solution so as to satisfy the Bragg diffraction condition. In the analysis of the TG signal, we assume that the solvent temperature rise rate, or the thermalization rate is given by a single exponential function with a rate constant of $k_{temp}$. This is the value we want to discuss in this paper. If the thermalization process is not a signal exponential but a multiple exponential process, the estimated time constant represents the weighted average of the heat released in each process.[9c]

The time profile of ISTS signal intensity, $I_{TS}(t)$, is expressed as[14]

$$I_{TS}(t) \propto |\delta\rho_{TS}(t)|^2 \quad (1)$$

where $\delta\rho_{TS}(t)$ is the difference from the equilibrium value of density due to the thermal expansion which is given by:

$$\delta\rho_{TS}(t) \propto \frac{1}{(k_{th}-d_a)^2+\omega_0^2}\left[\exp(-k_{th}t)+\left\{\left(\frac{k_{th}-d_a}{\omega_0}\right)\sin\omega_0 t-\cos\omega_0 t\right\}\exp(-d_a t)\right]$$
$$-\frac{1}{(k_{temp}-d_a)^2+\omega_0^2}\left[\exp(-k_{temp}t)+\left\{\left(\frac{k_{temp}-d_a}{\omega_0}\right)\sin\omega_0 t-\cos\omega_0 t\right\}\exp(-d_a t)\right] \quad (2)$$

where $\omega_0$ and $d_a$ are the acoustic frequency and the acoustic damping constant, respectively. A thermal decay rate coefficient, $k_{th}$, is given by $D_{th}(\omega_0/V)^2$, where $D_{th}$ and $V$ are thermal diffusion coefficient and the sound velocity of solvent, respectively. Approximately, $I_{TS}(t)$ oscillates in $2\pi/\omega_0$ period, because $\delta\rho_{TS}(t)$ is nearly proportional to $1-\cos(\omega_0 t)$ for $\omega_0 \gg d_a, k_{th}$.[15]

When the solute is excited by the pump pulse, ISTS is the main source of TG signal. However, when the absorbance of the solute is small, another contribution to the TG signal appears; that is, the impulsive simulated Brillouin scattering (ISBS) signal. Even for the non-resonant condition of the pump pulse, the electric field of interference fringe causes electrostrictive stress on solvent. This space modulation of stress induces acoustic stand wave, which is known to be the ISBS contribution.[16] In most cases of this study, absorbances of the azulene moiety are large enough to neglect this ISBS contribution. However, concentration of the samples in xenon cannot be large, because of the solubility problem, and the ISBS may appear weakly. This contribution, although it could be minor, should be taken into account in the analysis. The time profile of the density change by the ISBS component, $\delta\rho_{BS}(t)$, can be similarly derived as was done for $\delta\rho_{TS}(t)$[17] and given by:

$$\delta\rho_{BS}(t) \propto \frac{(\gamma-1)k_{th}q^2}{(k_{th}-d_a)^2+\omega_0^2}\left[\exp(-k_{th}t)+\left\{\left(\frac{(\gamma k_{th}-d_a)(k_{th}-d_a)+\omega_0^2}{(\gamma-1)k_{th}\omega_0}\right)\sin\omega_0 t - \cos\omega_0 t\right\}\exp(-d_a t)\right]$$

(3)

where γ denotes the specific heat ratio. Approximately, the ISBS contribution to acoustic TG signal, $|\delta\rho_{BS}(t)|^2$ oscillates in $\pi/\omega_0$ period, because $\delta\rho_{BS}(t)$ is approximately proportional to $\sin(\omega_0 t)$ for $\omega_0 \gg d_a, k_{th}$.[15]

### (2) Experimental Apparatus

Experimental setup was similar to the one in the previous study.[11,9g] Erbium doped fiber laser and Titanium:Sapphire regenerative amplifier system (Clark, CPA-2001) was the light source with 400 fs in pulse width and 775 nm in wavelength. Pump pulses, with 570 nm in wavelength were generated from a home-made non-linear optical parametric amplifier. In this study, four-fold amplification for the optical parametric amplifier was used instead of two-fold amplification that was used before. The pulse was split into two equivalent intensities. Probe pulse was delayed relative to pump pulses via an optical stage and injected to sample at the angle which satisfies Bragg's diffraction condition. The signal was detected by a Photomultiplier tube. When the solvent was CFC-113 or acetonitrile, the pump pulses were incident on the sample with 60 degree crossing angle, corresponding to $q$-value of $1.1 \times 10^7$ m$^{-1}$.[11] Pump pulse width was approximately 400nm. Probe pulse was a portion of light source i.e. 775 nm in wavelength. The pump and probe pulse intensities were approximately 0.5 μJ per shot. Spot sizes were ca. 300 and 250 μm for pump and probe pulses, respectively. For improving the signal to noise ratio in this study, we employed toggle-mode data acquisition: i.e. an optical chopper, introduced in one of the two pump beam paths, blocked the every other shot of pump pulses. The signals when the pump path was open and close were averaged separately. Then the "close" signal (offset) was subtracted from the "open" signal at each step of the optical delay stage (TG signal). The offset was mainly scattering of probe pulse on the cell surface. The relative intensity of offset to TG signal was up to 25 and 50 percent in organic solvents and xenon, respectively. Both the "open" and "close" signals were accumulated approximately for 15,000 shots at each delay time step for the cases of liquid solvents, and approximately for 30,000 shots for the case of xenon. The sample was enclosed in a quartz optical cell with 2 mm in optical path length and kept at the room temperature (ca. 25 °C). When the sample was xenon solution, the pump pulses were crossed at 180 degrees in the sample, corresponding to $q$-value of $2.2 \times 10^7$ m$^{-1}$.[9g] The probe pulse was the same color as the pump pulse (570 nm), and separated from the OPA output. The pulse width was adjusted to ca. 600 fs. The sample was enclosed in a high pressure optical cell with 2mm in optical path length. The detail of the high pressure cell and high pressure equipment is described elsewhere.[45,47] The pressure of the solution was kept within ±0.2 MPa range of the targeted value. The experiments for AzEzAn, AzPrAn, and AzHeAn were done at 35.8 MPa and 48.9 MPa, and those for AzCOCAn and AzCOCCOCAn were done at 28.8 MPa and 35.1 MPa. The cell temperature was controlled at 25±0.1 °C by water circulation using a thermostat. The sample solution was continuously stirred by a magnetic stirrer inside the high pressure cell during the measurements. The pump and probe pulse intensities were around 0.3 $\mu$J per shot. Spot sizes for pump and probe pulses were ca. 200 and 125 μm, respectively. Sample concentration was approximately 1.2 per 2 mm in optical density (ca. 0.1 mol/dm$^3$) at 570 nm for azulene. Other samples were saturated solutions which gave the optical density at 570 nm ranging from 0.2 to 1.0 per 2 mm in the organic solvents and approximately 0.1 per 2 mm in xenon, respectively.

**C. Computational methods**

Since the molecules treated here have a flexible alkyl-chain, at first we determined stable conformations by the semi-empirical MO method. For example, in the cases of 1-alkylzulene, five most stable conformations were estimated in the two-step procedures using Scigress Explorer package (version 7.5, Fujitsu). Firstly, stable conformations were generated from initial conformations with 24 evenly different angles by 15 degrees around each carbon-carbon single bond using CONFLEX MM2 method. Secondly, the heat of formation was calculated for each stable conformation after further optimization, using a MOPAC PM5 method. The number of stable conformers generated for bridged azulene-anthracenes, $N$, is listed in Table 1. In the cases of the bridged compounds, the chromophore distance $D$, the distance between the two edges of bridge part, was evaluated by the Canonical ensemble average of all stable conformers.

Vibrational normal mode frequencies and vectors were calculated by the means of density functional theory (DFT) using Gaussian03[18] for the five most stable conformers at B3LYP/6-31G(d) level after the structural optimization (four conformers for AzEtAn). The normal mode frequencies of the Canonical ensemble average of five conformations are given in supporting information (Figure S1, Tables S1 to S5, scaling factor: 0.9806).[19] The electronic energy difference was competitive to vibrational excess energy per single vibrational mode during VER process: approximately 10 kJ/mol in the most extreme case. The conformational difference altered vibrational frequencies by ca. 1 % at most for the normal mode above 1000 cm$^{-1}$, and by ca. 5 % for the normal mode above 100 cm$^{-1}$. Since anharmonisity calculation is computer demanding, only one conformer of the molecule proceeded through this calculation; namely, the anharmonic frequencies of azulene and 1-ethylazulene of the most stable conformation were calculated by the same DFT method after the structural optimization. Scigress Explorer was executed by a workstation. Gaussian 03 was executed by eight-processor parallel computing with 56 GB memory (73.6 GFLOPS, Hitachi SR11000).

## 3. Results

A typical acoustic TG signal of AzEtAn in CFC-113, acetonitrile and xenon are shown in figure 2(a)-(c). For each solvent, circles in the lower graph are experimental data. In order to extract the thermalization rate of the solvent ($k_{temp}$), we assumed that the thermalization kinetics is given by a single exponential kinetics and apply Eqs. (1) and (2) for the analysis. In practice, the TG signal was simulated by the following equation;

$$I_{TG}^{TS}(t) \equiv A\left|\delta\rho_{TS}(t)\right|^2 + Bt + C \qquad (4)$$

*A* is a scaling parameter, and empirical parameters B and C are also introduced to compensate offset and base line gradient of the signal, respectively. These parameters were optimized for each signal by non-linear least square method, except $k_{th}$. Since $k_{th} = D_{th}(\omega_0/V)^2$, the value was determined by the literature values of $D_{th}$ and $V$ using the optimized parameter $\omega_0$, respectively ($D_{th} = 1 \times 10^{-8}$ m$^2$/s and $V = 7 \times 10^2$ m/s[20] for CFC-113. $D_{th} = 1.27 \times 10^{-8}$ m$^2$/s and $V = 1.284 \times 10^3$ m/s for acetonitrile. $D_{th} = 5.30 \times 10^{-8}$, $5.41 \times 10^{-8}$, $5.41 \times 10^{-8}$, and $9.41 \times 10^{-8}$ m$^2$/s and $V = 4.3004 \times 10^2$, $5.2411 \times 10^2$, $5.2410 \times 10^2$, $5.2410 \times 10^2$ m/s for xenon at 28.8, 35.1, 35.8, and 48.9 MPa, respectively.) The optimized curves are also shown as solid lines in Fig. 2. The upper graphs show residuals of the fitting, and the vertical scale of the residual plot is normalized by the peak intensity of the TG signal. As is shown in the figure, Eq. (4) reproduced the signal well in acetonitrile and CFC-113.

Due to the low solubility of the samples in xenon, however, the relative intensity of the TG signal to the scattering of the probe light was about twice smaller than those in organic solutions. This weaker signal may cause a technical problem. When we fit the observed TG signal by Eq. (4), we found the residuals of the fitting oscillate sinusoidally as is shown in Fig. 2(c) (open square in upper graph). We considered that the impulsive stimulated Brillouin scattering from xenon was contaminated in the acoustic signal. Therefore we fit the observed signal by a function adding a term of $\delta\rho_{BS}(t)$ inside the square bracket of Eq. (4), i.e.:

$$I_{TG}^{total}(t) \equiv A\left|\delta\rho_{TS}(t) + E\delta\rho_{BS}(t)\right|^2 + Bt + C \tag{5}$$

where $E$ is an adjustable parameter. The result of the fit for Eq. (5) is shown in Fig.2 (c) (solid line in lower graph). No sinusoidal oscillation is found in the residuals, and the signal was well reproduced by the model.

The thermalization rates obtained by the fit are summarized in Table 1. The results for the xenon solutions are the average of the results at two different pressures, since we did not detect noticeable pressure dependence. Experimental errors were evaluated as the sum of the mean square root of accidental error for least square estimations and systematic error of each experimental run, with 68.26 % (first standard deviation) confidence levels.[21] The systematic error of $\omega_0$ and $d_a$ did not affect $k_{temp}$ values significantly.

## 4. Discussion

**A. General features of the thermalization**

As is shown in the Table 1, the thermalization rates $k_{temp}$ from bridge azulene-anthracenes are larger than that of azulene irrespective of solvent. The previous results of the TA study also indicated that the CET rates of bridge azulene-anthracenes were larger than that of azulene. In this section, we will discuss a general origin of the acceleration of the thermalization rates of bridged compounds, before the detailed investigation of chain length dependence. As is shown in Table 2, an addition of 1-alkyl group to azulene accelerates the thermalization rates significantly in CFC-113 and acetonitrile. In addition, there is no alkyl chain length dependence between two 1-alkylazulenes in acetonitrile solution. Therefore, the acceleration of the thermalization rate is considered to be mainly induced by adding an alkyl-chain to the azulenyl group.[22]

Why is the local energy dissipation from azulenyl group accelerated? According to a non-equilibrium classical MD simulation of azulene in carbon dioxide, the three lowest frequency modes dissipates the excess energy in ca. 200 fs, while the total VER takes around 20 ps.[23] The results indicate that IVR in azulene is incomplete during the total energy dissipation process and that the IVR is the rate determining step of the VER process in azulene. Therefore, it is rational to attribute the acceleration effect in 1-alkylazulenes to the IVR stimulation. In condensed media, the IVR process is composed of vibrational state transitions inside the solute molecule under the fluctuating force from the solvent. Landau-Teller formula[24] says that, in principle, the transition probability is inversely proportional to the energy gap and to the square of coupling constant between the initial and final states. Therefore, IVR rate depends on the density of vibrational states through the former factor and higher order Hessians through the latter factor. In order to evaluate the former factor, the densities of states are calculated by a harmonic state count method using the results of the vibrational modes by the DFT calculations. Figure 3 compares the densities of the state of azulene and 1-ethylazulene against the vibrational energy. The density of state of 1-ethylazulene is 10 and 100 times larger than that of azulene at 500 and 2500 cm$^{-1}$, respectively. This clearly indicates that the possibility of the acceleration of the IVR by adding alkyl-chain to azulene.

Since an exact evaluation of the coupling constant between the initial and final states (the second factor) is a difficult task, we tried to extract the characteristics by evaluating the second-order anharmonic coefficients between each pair of vibrational normal modes. This trial is based on a consideration that the low-order coupling drives IVR kinetics generally.[25] We calculated the anharmonic coefficients by the second-order perturbation theory implemented into Gaussian03. In this model, the vibrational energy $E_V$ is given by

$$E_V = \sum_r{}' \omega_r (\upsilon_r + \tfrac{1}{2}) + \sum_{r \geq s}{}' \chi_{rs} (\upsilon_r + \tfrac{1}{2})(\upsilon_s + \tfrac{1}{2}) \tag{6}$$

where $\omega_r$ is the $r$th harmonic vibrational frequency, and $\chi_{rs}$ is the vibrational anharmonic constants, respectively. We calculated the average of $\chi_{rs}$, i.e., $\tfrac{1}{M}\sum_s \chi_{rs}$ value for each normal mode of azulene and 1-ethylazulene, where $M$ is the number of the normal modes. This value may be an indicator of the coupling strength of a normal mode to other normal modes. Figure 4 shows the results against the each normal mode frequency. The distributions of the anharmonicity are quite similar to each other, except three C-H stretching modes of AzEt: the 61st, 62nd and 66th mode in the ascending order of frequency. These normal modes have more than ten times larger anharmonic coefficients on average than the other C-H stretching modes. The vibrational modes which couple to these modes are other

C-H stretching modes and out-of-plane C-H bending modes from 350 to 500 cm$^{-1}$. Because of this strong coupling, these three normal modes could mediate intramoleculer energy transfer from high frequency modes to low frequency gateways.

We mention here the effect of the chain-length and chain-species. In general, the longer chain-length compounds of the same kind of chain shows the faster relaxation. Further, in comparison of the results for 1-alkylazulenes with those bridged-compounds, the thermalization rates from the bridged compounds are slower. An addition of an anthracen-9-yl group to a 1-alkylazulehne chain decelerates the thermalization. This phenomenon can be explained by assuming that the local energy dissipation rate of azulen-1-yl group is significantly faster than that of anthracen-9-yl group. Zalesskaya measured VER rate of anthracene in the electronic ground state after the infrared multiphoton excitation.[26] They determined the VER rate in gaseous phase, and estimated the buffer gas pressure dependence of the energy dissipation rate.[27] Their estimated value for anthracence was generally much slower than that for azulene. Although they did not measured the effect of its 9-alkylation, Phillips reported that 9-methy substitution does not differ VER rates of $S_1$ anthracene.[28] Further according to Zalesskaya, VER rates of anthracene in the electronic ground state are almost equivalent to that in the $S_1$ state.[26] Therefore, the VER kinetics of 9-alkylanthracenes in the ground state is expected to be much slower than those of 1-alkylazulene.

**B. Intramolecular Vibrational Energy Redistribution**

In order to examine the intermolecular energy transfer during the IVR process, our solvent thermalization rates are compared with the solute relaxation rates obtained from the previous TA experiments. The TA data are available for AzPrAn, AzHeAn, AzCOCAn and AzCOCCOCAn in xenon solution and for AzEtAn and AzPrAn in CFC-113.[10a] As is mentioned in our previous paper,[9g] the energy dissipation rate in xenon should be identical to the solvent thermalization rate, since in xenon there is no intermolecular V-V energy transfer which produces a delay of the solvent thermalization rate to the VER rate of the solute. If the vibrational excess energy of the solute dissipated to the solvent during the "IVR" process, the thermalization rate may be faster than the "CET" rate determined by the TA measurement due to the contribution of the faster energy dissipation. From this point of view, it is said that the IVR component in xenon observed by TA experiment does not involve energy dissipation to environment, since the solvent thermalization rates of four compounds in xenon are identical to corresponding CET rate of the TA experiments within experimental error. We are not sure of the reason of the slower thermalization rate of AzPrAn, which suggests that the slower energy dissipation process may exist which cannot be detected by the TA study.

On the other hand, in CFC-113 the relative speed of the thermalization rate to the CET rate is dependent on the solute. According to the study by the TA, CET rates are independent from solute species within the experimental error, as long as azulene, AzEtAn and AzPrAn are concerned. Therefore, it is a reasonable assumption that the CET rate for AzHeAn is equivalent to those of three solute compounds mentioned above, namely ca. 50 ns$^{-1}$. Under this assumption, AzHeAn seems to thermalize faster than the CET rate in CFC-113. The larger $k_{temp}$ than the assumed CET rate means that the "IVR" process involves the solute-solvent interactions. According to the TA experiments, the excess vibrational energy localized in the azulenyl group transfers through the bridge part to the anthracenyl group to achieve the statistical equilibrium distribution at the rate of around 200 ns$^{-1}$. A portion of the vibrational energy current along the bridge structure during "IVR" may dissipate to the environment.

Why is the energy dissipation in CFC-113 for the bridged-compounds outstandingly efficient? In order to estimate the physical background, we tested the correlation of the thermalization rate with several physical parameters of the solute molecules which may be related to the VER. Since the thermalization rate in CFC-113 generally becomes faster with the longer chain, we selected the following three parameters: the number of density of vibrational states less than 500 cm$^{-1}$ ($N$), the number of ethylene groups along the bridge ($n_{Me}$), and the average distance between azulenyl and anthracenyl groups ($D$). The first parameter tests the V-V and V-T energy transfer efficiency in the lower region, and second one tests the specific interaction of $CH_2$ group with the solvent, and the last one tests the accessibility of the solvent to the solute. Figures 5(a)-(c) shows the correlation. In all cases, the thermalization rate increases as the parameter increases. The correlation coefficients ($r$) show that $n_{Me}$ and $D$ are equally good scaling parameters ($r=0.98$) relative to $N$ ($r=0.83$). Since the number of the low-frequency vibrational modes less than 500 cm$^{-1}$ of CFC-113 (11, see Table S6) is larger than that of acetonitrile (1,380 cm$^{-1}$)[29] and xenon (0), an increase of the number of the low-frequency modes of the bridged-compounds may be correlated with the thermalization rate if a simple V-V energy transfer is effective. According to Fig. 5(a), however, an increase of the vibrational density states of the lower-frequency modes does not significantly accelerate the thermalization rate.

Although the mechanism of the energy dissipation from the alkyl-chain in CFC-113 is not clear at present, we will mention plausible reasons for the acceleration based on the successful scaling parameters $n_{Me}$ and $D$ for CFC-113. The correlation with $n_{Me}$ suggests that the coupling of the methylene group to the solvent affects the energy dissipation from the bridged region. Since CFC-113 does not have the vibrational frequency above 1212 cm$^{-1}$(see Table S6)[30], the direct resonant coupling with the C-H stretching mode (around 3000 cm$^{-1}$) or the H-C-H bending mode group (around 1400cm$^{-1}$) is impossible. If the V-V energy transfer is effective, the overtone or the combination mode of CFC-113 or the lower frequency skeleton vibration of the solute may be a candidate. On the other hand, the distance between the chromophores ($D$) is related to the accessibility of the solvent molecules to the bridge part of the solute molecule. The shorter $D$ is, the more difficult the access of solvent to the solute's bridge part must become. Therefore, it is also probable that the collisional interaction between the chain part and the solvent is important. In general, the collisional interaction mediates V-T and V-R energy transfer, while the longer range interactions such as the dipole-dipole interaction intermediate the V-V energy transfer. Since the energy accepting state must be specific to CFC-113, the accepting mode may not be translation but rotation, because translational phonon modes are similar to each other for typical organic solvents, in general. The larger inertia of momentum of CFC-113 may be related to the feature distinguishable from acetonitrile. To see this point in more detail, the molecular dynamics simulation of realistic model molecules is now under investigation.

## 5. Conclusions

In this work, we applied the TG acoustic spectroscopy to the vibrational energy relaxation process of the bridged azulene-anthracenes in order to test the possibility of the energy dissipation from the solute to the solvent during the "IVR" process assigned by the TA measurements on the same systems. According to the experimental results in xenon, the solvent thermalization rate was identical to the CET rate determined by the TA measurement, and the energy dissipation to the solvent during the "IVR" process was not detected for the solution of xenon. On the other hand, in CFC-113, the relative speed of the thermalization rate to the CET rate was dependent on the solute species, and especially faster solvent thermalization rate of AzHeAn was predicted. The result suggests that the faster energy dissipation process than the CET process; i.e., the energy dissipation must occur before IVR process completes. However, "IVR" rates are independent from solvent species according to the TA experiments. In other words, the solute-solvent interactions must control energy transfer paths but take no effect on energy transfer rates.

The underlining mechanism of the faster energy dissipation in CFC-113 is still not unclear. In order to see more detail on the effect of the chain part, we are now planning to study the thermlization process with different chain lengths of 1-alkylazulenes. Since 1-alkylazulenes are more easily soluble to various solvents than bridged azulene-anthracenes, the effects of the V-V coupling and the solvent rotational modes may be tested by using various solvents from xenon to complex ones. These experimental studies together with the molecular dynamics simulations will bring more detailed insights into the VER mechanism of this kind of materials.

**ACKNOWLEDGMENT**

We are owing to many helpful advices and comments, especially in organic synthesis (Appendix) and Calculation (Section 3) sections. This work was partially financed by Grants-in-Aid for Scientific Research from JSPS (Nos. 16350010 and 19350010). H. F. was supported by research fellowship of Global COE program, International Center for Integrated Research and Advanced Education in Material Science, Kyoto University, Japan.

**Supporting Information Available.**

The details of the synthetic procedure of bridged azulene-anthracences, typical optimized structures of bridged azulene-anthracences by DFT calculations (B3LYP/6-31G$^*$), the vibrational frequencies of normal modes of bridged azulene-anthracences for the optimized structure by the DFT calculation, and the vibrational frequencies of the normal modes of CFC-113.

**Table 1**

Thermalization rate constants on bridged azulene-anthracenes in various solvent fluids at 25.0 °C.

| Probe Molecule | Acetonitrile | Xenon | | CFC-113[a] | | $k_{IVR}$ / ns$^{-1}$ | $N$[g] |
| --- | --- | --- | --- | --- | --- | --- | --- |
| | $k_{temp}$ / ns$^{-1}$ | $k_{temp}$ / ns$^{-1}$ | $k_{CET}$ / ns$^{-1}$[b] | $k_{temp}$ / ns$^{-1}$ | $k_{CET}$ / ns$^{-1}$[b] | | |
| AzEtAn | 45±3 | 17±1 | - | 39±1 | 43±4 | 270±29 | 14 |
| AzPrAn | 45±4 | 19±1 | 29±2[c] | 48±7 | 50±5 | 270±22 | 41 |
| AzHeAn | 50±2 | 25±2 | 25±3[d] | 76±16 | - | 238±34 | 272 |
| AzCOCAn | 53±2 | 30±9 | 25±2[e] | 36±3 | - | 200±16 | 28 |
| AzCOCCOCAn | 59±1 | 25±2 | 29±3[f] | 65±13 | - | 196±15 | 142 |

[a] 1,1,2-trichloro-1,2,2-triflutoroetane

[b] Ref. 10(a).

[c] 48.9 MPa.

[d] 35.8 MPa

[e] 28.8 MPa.

[f] 35.1 MPa

[g] The number of conformations found by PM5 calculation (see text).

**Table 2**

Thermalization rate constants on azulene and 1-alkylazulene`s in acetonitile and CFC-113[a,b]

| Probe Molecules | CFC-113 | Acetonitrile |
|---|---|---|
| Azulene | 29±1 | 35±2 |
| AzEt | 140±24 | 59±1 |
| AzDo | 221±63 | 59±7 |

[a] Ref. 11.
[b] unit: ns$^{-1}$

**Scheme 1.** Molecular structures of bridged azulene-anthracences.

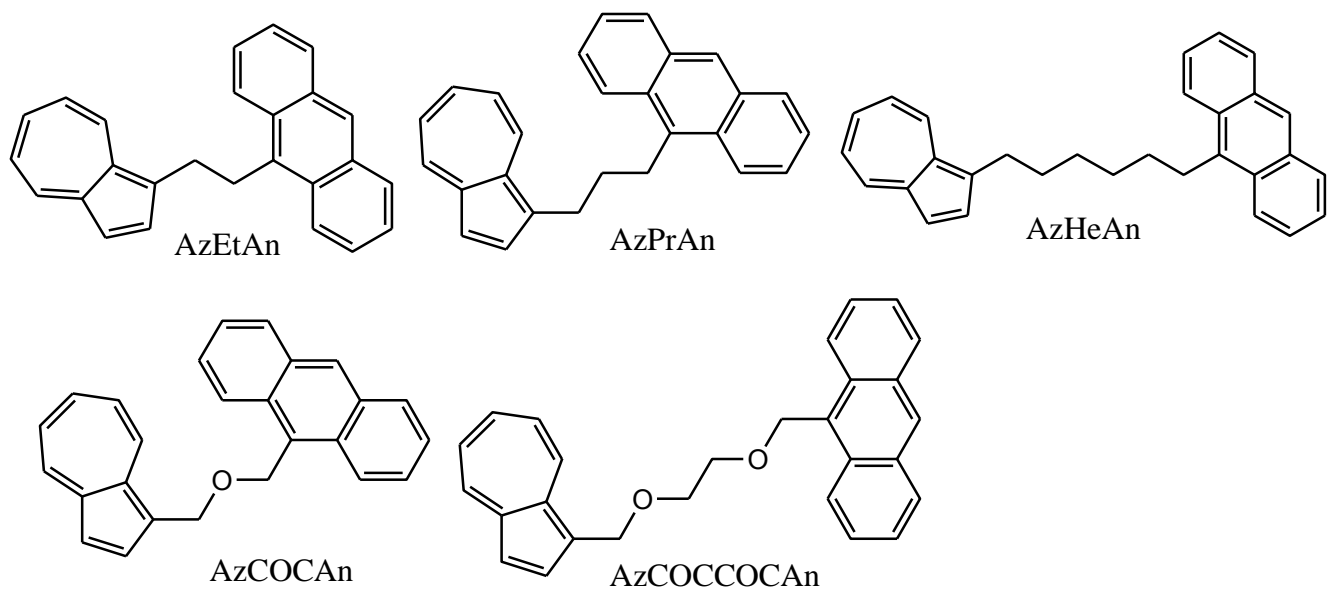

**Figure 1.**

Experimental apparatus aligned for supercritical xenon solutions. LBO is Lithium Triborate (LiB$_3$O$_5$) crystal, in which the major portion of laser output generates second harmonic, the pump pulse of a home-made optical parametric amplifier (OPA). S is a sapphire parallel plate generating white light, the seed pulse of OPA from the minor portion of the laser output. BBO1 is $\beta$-bariumborate ($\beta$-BaB$_2$O$_4$, BBO) crystal for the first and second amplifications of OPA. BBO2 is another BBO crystal used for the third and fourth amplifications of OPA. Two pump pulses are incident on the sample in a counter propagating alignment to optical window configurations of high pressure cell (HP Cell). P is Gran-Teller polarizer, PM photo-multiplier tube, respectively.

**Figure 2.**

Least square fitting is demonstrated to typical acoustic transient grating (TG) signals in each solvent. (a) 1,1,2-trichloro-1,2,2-trifluoroethane (CFC-113) solution of 9-(2-(azulen-1-yl)ethyl)anthracene (AzEtAn). Lower graph shows an acoustic signal (circle) and theoretical curve for ISTS (Eq. (4), solid line). Upper graph displays residuals of the fitting (circle). Dotted line indicates zero. (b) AzEtAn data in acetonitrile (circle) and ISTS curve (solid line). (c) AzEtAn data in xenon at 35.8 MPa (circle). ISTS curve and residuals are dotted line and open squares, respectively. The residual of the fit to Eq. (4) in xenon oscillated at two-fold frequency of the acoustic TG signal, while residuals in other solvents showed random distribution (see the text). ISBS term is necessary for better reproduction of data in xenon (simulation by Eq. (5): solid line, residuals: filled circle).

**Figure 3.**

Density of the intramolecular vibrational states for azulene (lower curve) and 1-ethylazulene (upper curve). The harmonic state count method was applied to the normal mode frequencies estimated by the DFT method (B3LYP/6-31G$^*$) using Gaussian 03.

**Figure 4.** Square average of the anharmonic coefficients of the normal mode for azulene (circle) and 1-ethylazulene (AzEt, triangle). The second order perturbation theory (B3LYP/6-31G$^*$) was used (Gaussian 03). Three illustrations are corresponding normal mode vibrations of AzEt to pointed data. Black and green lines are chemical bonds and atomic motions, respectively.

**Figure 5.**

Translational temperature rise rate constants $k_{temp}$ of solvents (filled makers) determined by the transient grating (TG) experiment are plotted against (a) the number of vibrational states ($N$), (b) the number of methylene group in the chain part ($n_{Me}$), and (c) the average chromosphere distance $D$ of solute molecule at 650 K (see the text). Circle, triangle and square represent 1,1,2-trichloro-1,2,2-trifluoroethane (CFC-113), acetonitrile and xenon, respectively. Vertical bars indicate experimental error. Horizontal bars in (c) indicate temperature dependence of $D$ within 300 to 1000 K. For CFC-113 and xenon solution, the collision energy transfer (CET) rates determined by transient absorption (TA) are also quoted from literature (open makers). All axes are in a log scale.

Figure 1  H. Fujiwara et al.

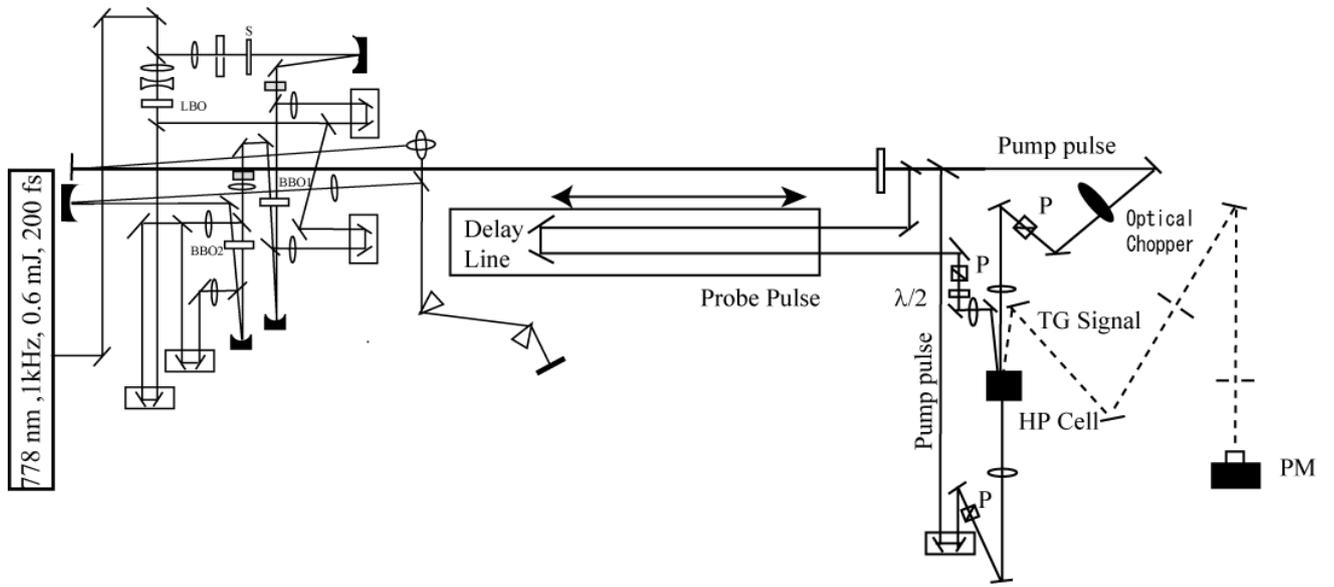

Figure 2 H. Fujiwara et al.

(a)

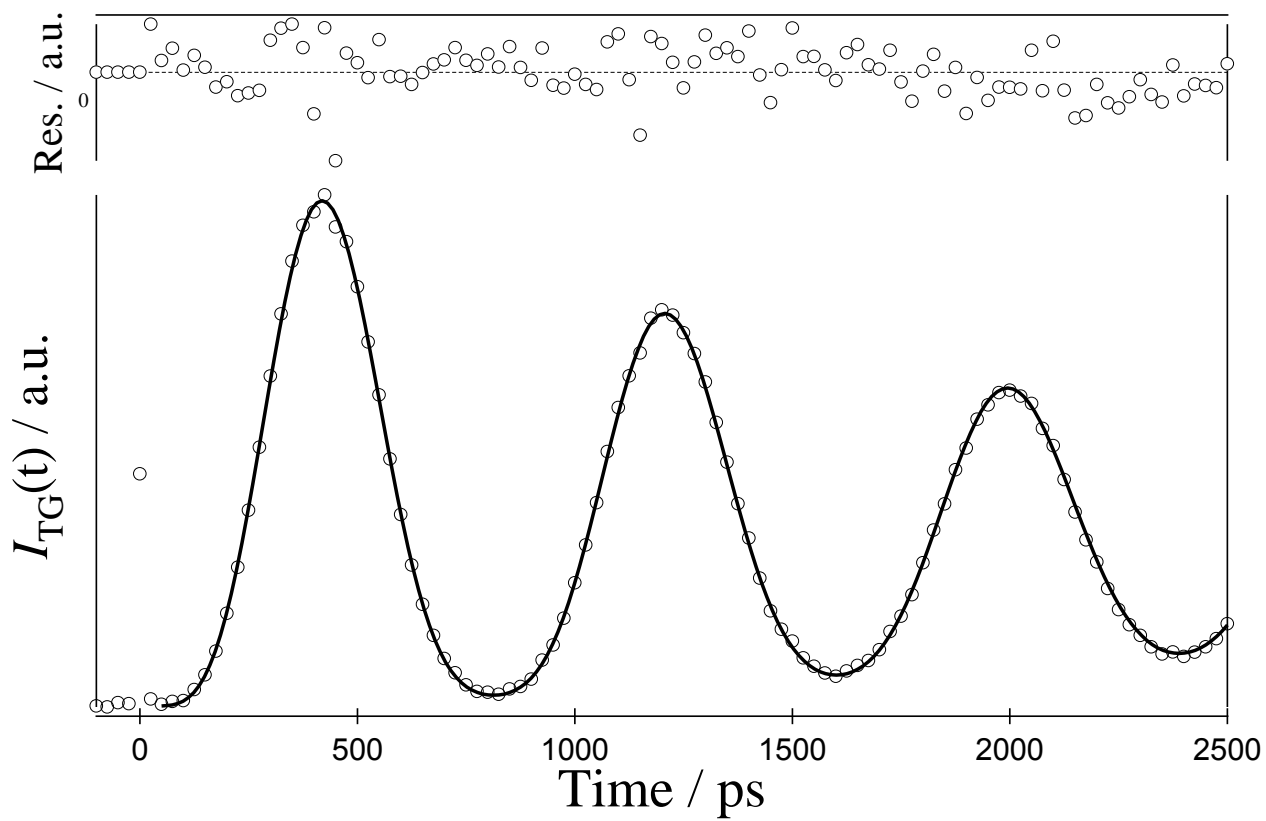

(b)

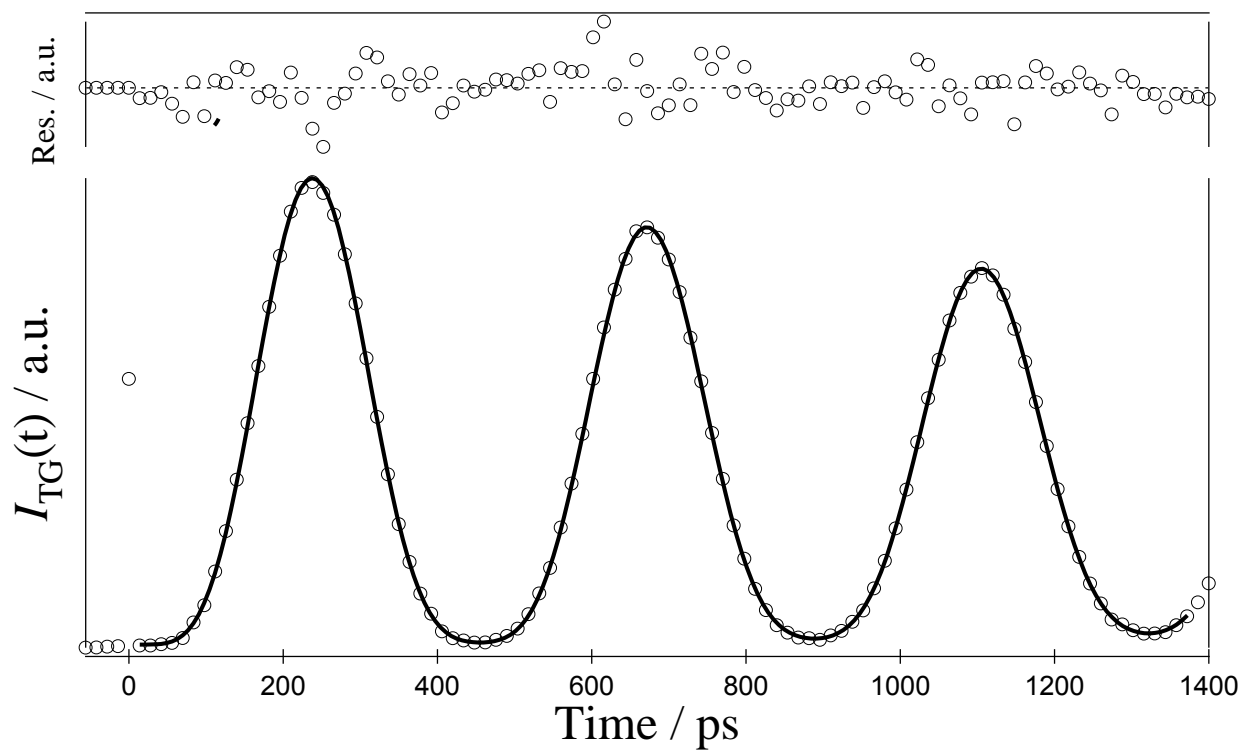

(c)

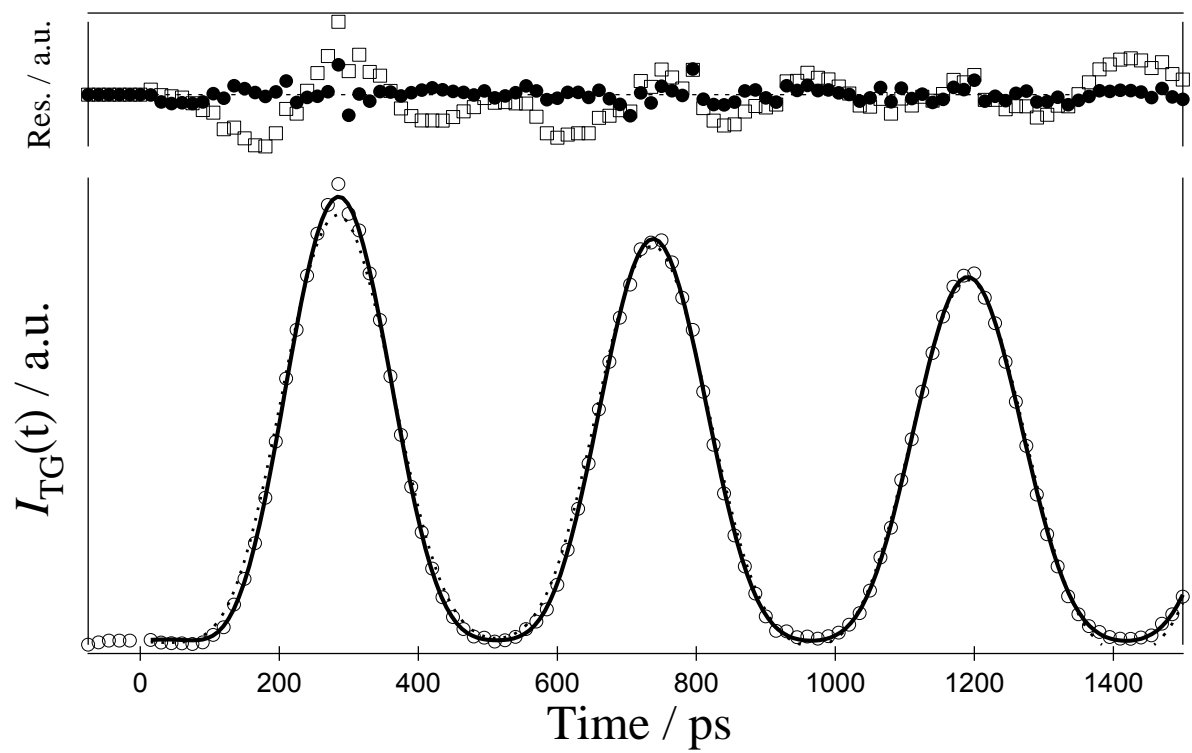

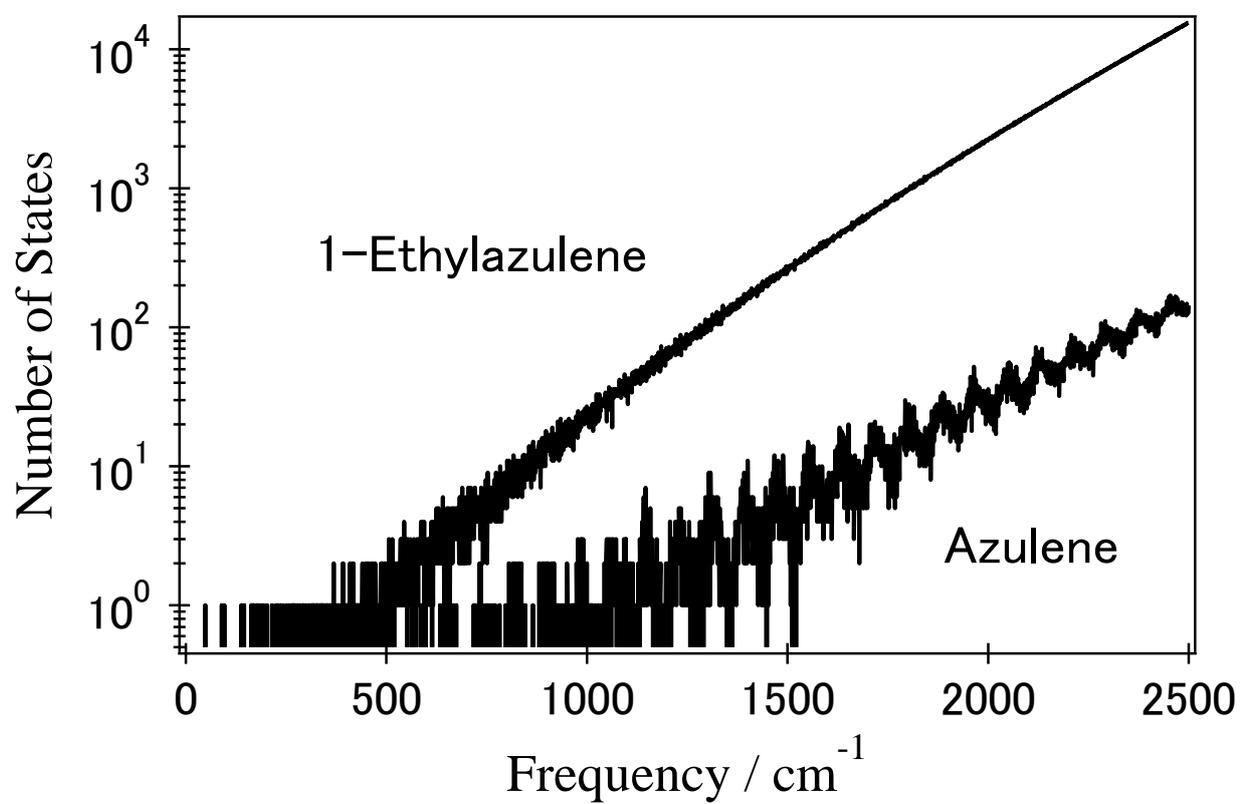

Figure 4  H. Fujiwara et al.

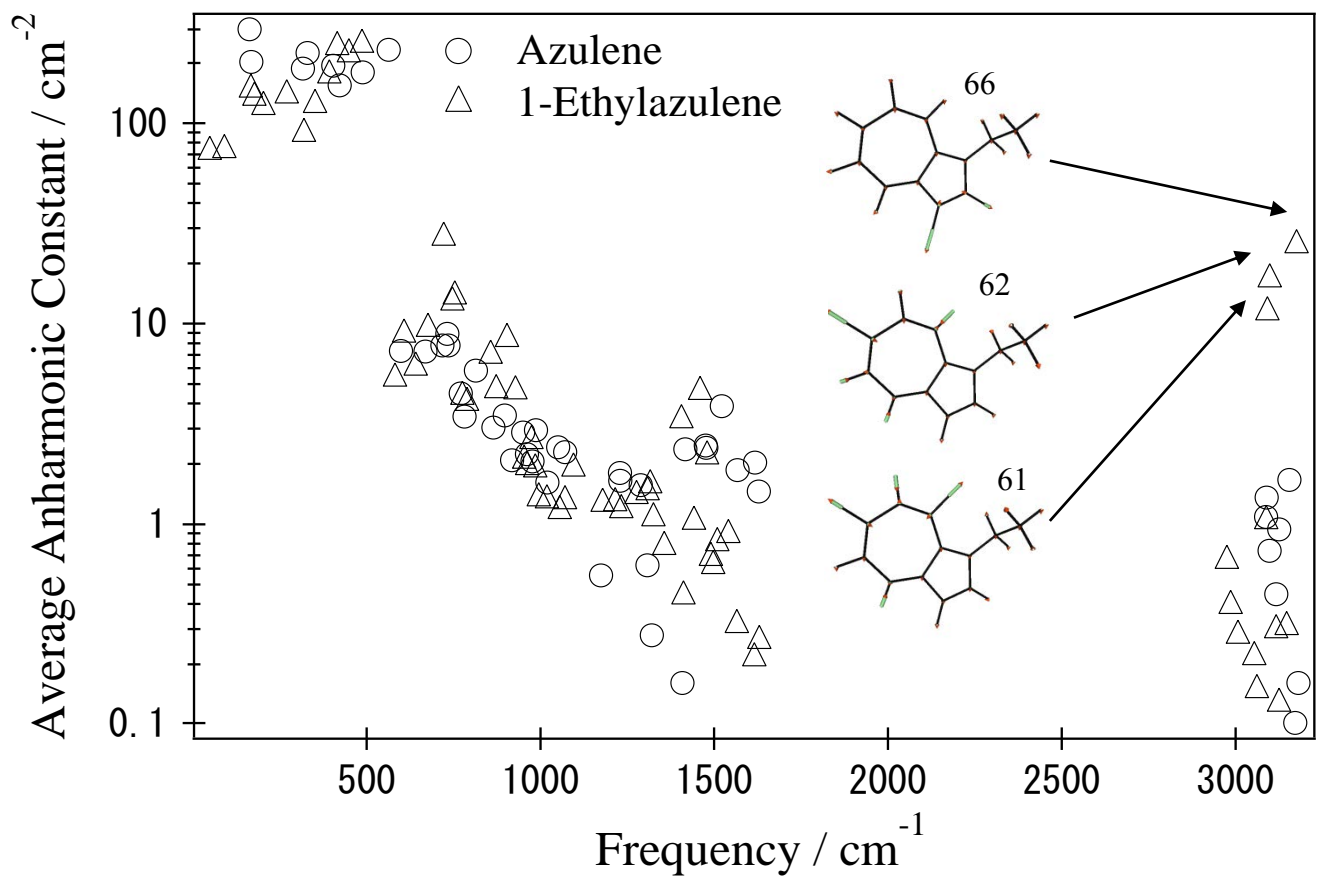

Figure 5 H. Fujiwara et al.

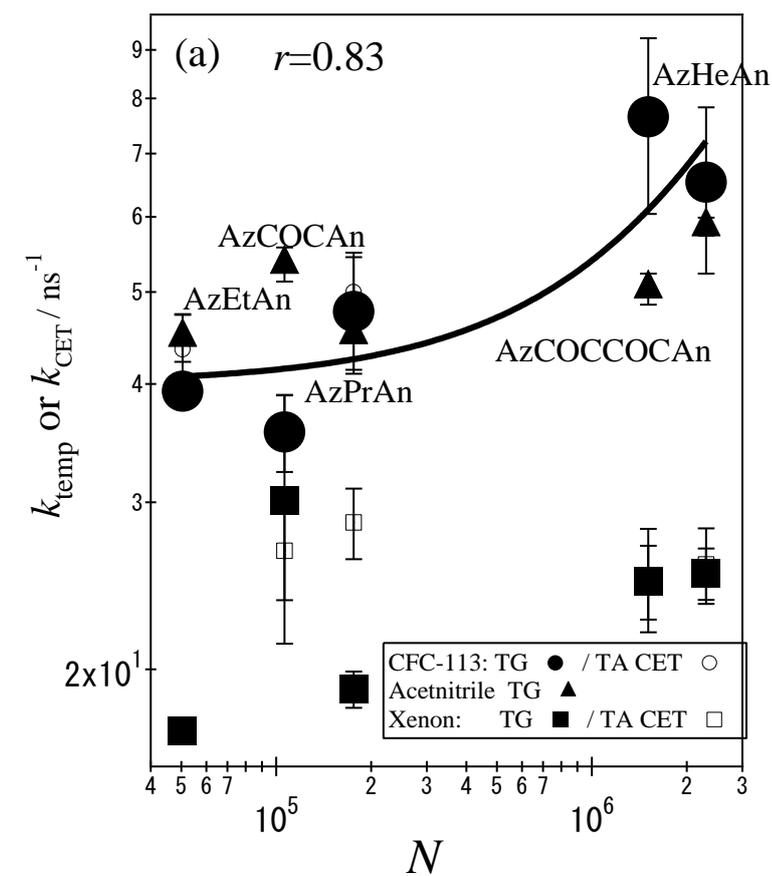

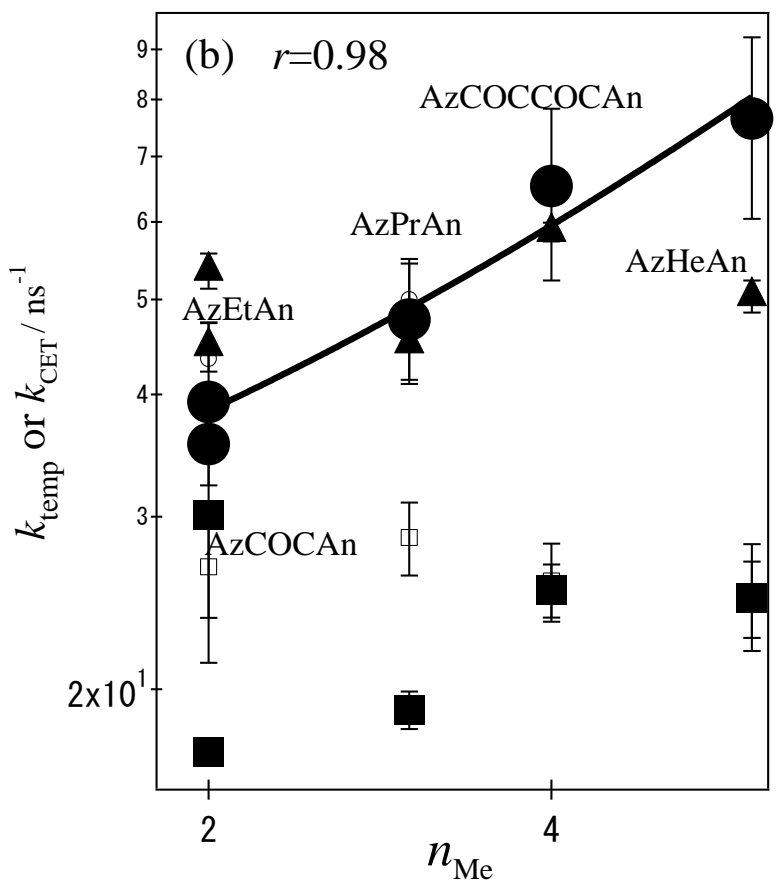

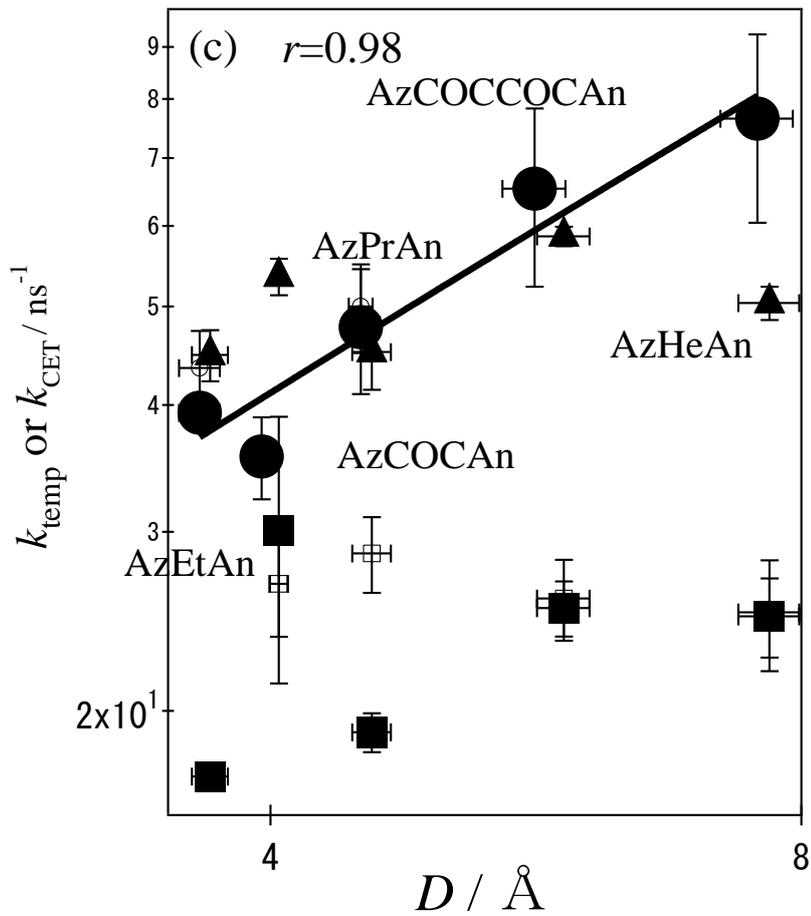

# SUPPORTING INFORMATION

Synthesis

Synthetic path ways to bridged azulene-anthracene compounds.

**Scheme S1**

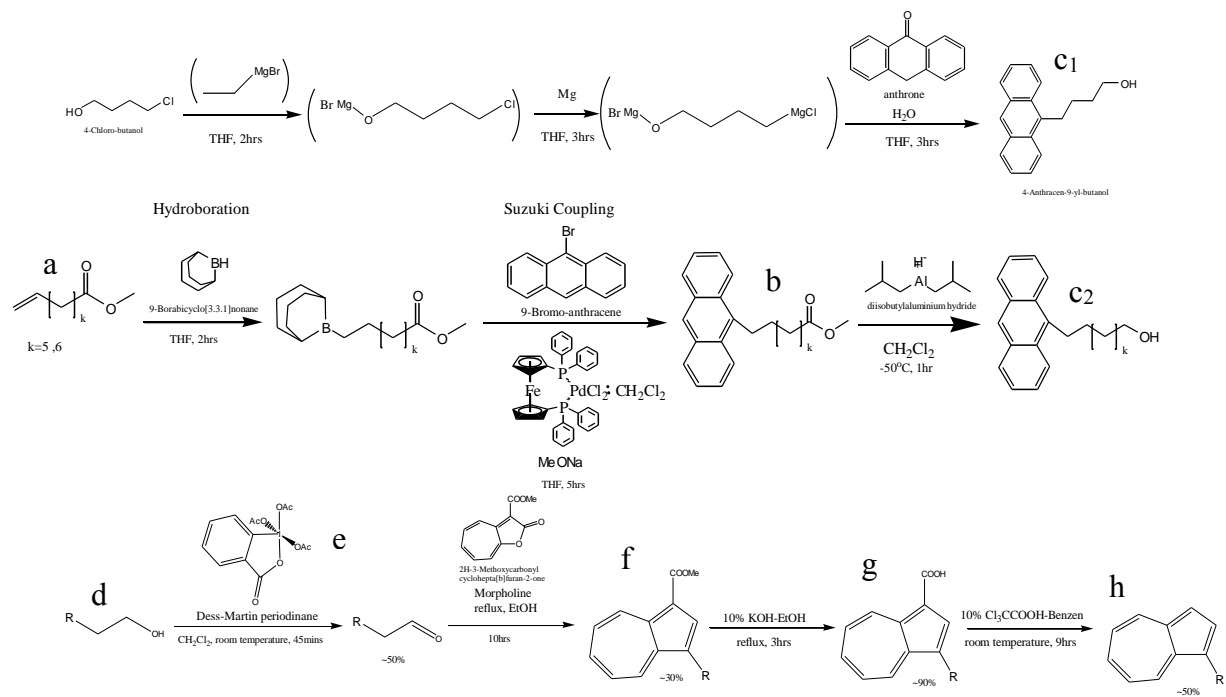

Alkyl bridged azulene-anthracenes were synthesized according to previously reported methods.

### 5-(Anthracen-9-yl)pentanoic Acid Methyl Ester and 8-(Anthracen-9-yl)octanoic Acid Methyl Ester: General Procedure[12(c)]

9-Borabicylco[3.3.1]nonane was added to a THF solution of compound (a) (0.5 mol/l, 1 eq. mol), and the solution was stirred at room temperature. After 2 h, 9-bromoanthracene (0.9 eq. mol), [1,1'-bis(diphenylphosphino)ferrocene]palladium(II) dichloride (0.027 eq. mol), and a THF solution of sodium methoxide (1 mol/l, 0.27 eq. mol) were added in this order, and the resulting solution was refluxed for 5 h. The mixture was then cooled to room temperature, and filtered through Cellite.

Brine was added to the filtrate, and the mixture was extracted with toluene. The combined organic extracts were dried, evaporated, and subjected to silica gel column chromatography ($CH_2Cl_2$). Compound (b) was obtained as a pale yellow solid by recrystallization from methanol (ca. 50 %).

### 5-(Anthracen-9-yl)pentanol and 8-(Anthracen-9-yl)octanol: General Procedure[12(a)]

Diisobutylalminium hydride (2.2 eq. mol) was added to a $CH_2Cl_2$ solution of 5-(anthracen-9-yl)alkanoic acid methyl ester (0.2 mol/l, 1 eq. mol) at -50 °C, and the resulting solution was stirred for 1 h and then quenched by a small amount of saturated ammonium chloride aqueous solution. Cellite and ether were added to the reaction mixture, which was further stirred for 1 h. The organic layer was dried over sodium sulphate and concentrated in vacuo. The solid residue was subjected to silica gel column chromatography ($CH_2Cl_2$/ethyl acetate=10/1). Compound ($c_2$) was obtained as a white solid by reprecipitation from hexane (ca. 95%).

### 4-(Anthracen-9-yl)butanol

A THF solution of ethyl bromide (22 mmol/l, 2.2 eq. mol) was added dropwise to a flask containing magnesium (2 eq. mol). 1-Chlorobutanol (2 eq. mol) was added dropwise to the reaction mixture at 0 °C. After stirring for 30 min at room temperature, magnesium (2 eq. mol) was added into the flask, which was then gradually heated to 60 °C. After magnesium was consumed completely, the reaction mixture was cooled down to 0 °C, at which temperature anthrone (1 eq. mol) was added in portion. The resulting solution was refluxed for 3 h and then hydrolyzed by a saturated ammonium chloride aqueous solution (equal volume) at 0 °C. After stirring for 30 min, the organic phase was separated, dried over magnesium sulphate, concentrated and stood in a refrigerator. The liquid phase was separated, evaporated and subjected to silica gel column chromatography (benzene). Compound ($c_1$) was isolated as a pale yellow solid (ca. 5%).

### Anthracene-9-ylalkyl Aldehyde: General Procedure[12(d)]

A $CH_2Cl_2$ solution of Dess-Martin periodinane (15 w/w%, 1.1 eq. mol) was added to a $CH_2Cl_2$ solution of compound (d) (0.18 mol/l, 1.0 eq. mol). After 45-min of stirring, diethyl ether (2.8-fold volume) and a sodium hydroxide aqueous solution (1.3 mol/l, 4.6-fold volume of $CH_2Cl_2$) were added, and the resulting mixture was stirred for 30 min. The organic layer was separated, and the aqueous phase was extracted with $CH_2Cl_2$. The combined organic extracts were dried over magnesium sulphate, evaporated and subjected to silica gel column chromatography ($CH_2Cl_2$). Compound (e) was isolated as a white solid.

### 3-((Anthracen-9-yl)alkyl)azulene-1-carboxylic acid methyl ester: General Procedure[12(c-e)]

3-(Methoxycarbonyl)-2H-cyclohepta[b]furan-2-one (1g, 1eq. mol) was dissolved in EtOH (100 ml) at 50 °C. Morpholine (5 eq. mol) and compound (e) (1 eq. mol) were added, and the resulting mixture was refluxed for 12 h under a nitrogen atmosphere. After cooling to room temperature, the solution was evaporated, and the solid residue was subjected to silica gel column chromatography ($CH_2Cl_2$). Compound (f) was isolated as a purple solid.

### 3-((Anthracen-9-yl)alkyl)azulene-1-carboxylic acid: General Procedure[12(b-d)]

Compound (f) was dissolved in an alkaline ethanol solution (15 w/w% potassium hydroxide) and heated at 70 °C for 6 h under a nitrogen atmosphere. After cooling to room temperature, the solution was poured into water, neutralized to pH 5 using hydrochloric acid, and stirred for 30 min. Reddish-purple precipitates were filtered, washed by water and dried in vacuo for one day. The crude product was used for the following reaction without further purification.

### 9-((Azulen-1-yl)alkyl)anthracene: General Procedure[12(c-e)]

The carboxylic acid (compound (g)) was added to a benzene solution of trichloroacetic acid (15 w/w%), and the resulting solution was stirred at room temperature under a nitrogen atmosphere in the dark. After 12 h, the solution was poured into water, and the benzene layer was separated, washed with five-fold brine/sodium hydrogen carbonate-brine solutions, dried over sodium sulphate, and evaporated. The target compound (h) was purified twice by silica gel column chromatography ($CH_2Cl_2$). All the work-up manipulations were conducted at room temperature. AzEtAn and AzHeAn were obtained as violet solids. AzPrAn was obtained as a dark violet semi-solid.

**9-(2-(Azulen-1-yl)ethyl)anthracene**: Purple powder; $^1$H NMR (CD$_3$COCD$_3$, 400MHz) δ = 3.43 (t, 2H, *J* = 8.2 Hz), 3.96 (t, 2H, *J* = 8.4 Hz), 7.00 (t, 1H, *J* = 9.6 Hz), 7.01 (t, 1H, *J* = 9.6 Hz), 7.27 (d, 1H, *J* = 3.6 Hz), 7.39 (m, 4H), 7.48 (t, 1H, *J* = 10.2 Hz), 7.90 (d, 1H, *J* = 3.6 Hz), 7.96 (d, 2H, *J* = 8.0 Hz), 8.21 (d, 1H, *J* = 8.4 Hz), 8.23 (d, 1H, *J* = 8.0 Hz), 8.29 (d, 2H, *J* = 8.4 Hz), 8.36 (s,1H).

**9-(3-(Azulen-1-yl)propyl)anthracene**: Purple semi-powder; $^1$H NMR (CD$_3$COCD$_3$, 400MHz) δ = 2.28 (quintet, 2H, *J* = 4.0 Hz), 3.44 (t, 2H, *J* = 7.4 Hz), 3.72 (t, 2H, *J* = 8.4 Hz), 7.14 (t, 2H, *J* = 10.0 Hz), 7.44 (m, 5H), 7.61 (t, 1H, *J* = 9.8 Hz), 7.94 (d, 1H, *J* = 3.6 Hz), 8.04 (d, 2H, *J* = 7.2 Hz), 8.17 (d, 2H, *J* = 9.6 Hz), 8.36 (d, 1H, *J* = 9.2 Hz), 8.41 (s, 1H), 8.47 (d, 1H, *J* = 10.0 Hz); MS (MALDI-TOF) m/z 344 (M+).

**9-(6-(Azulen-1-yl)hexyl)anthracene**: Purple powder; $^1$H NMR (CD$_3$COCD$_3$, 400MHz) δ = 1.52 (quintet, 2H, *J* = 7.8 Hz), 1.67 (quintet, 2H, *J* = 7.7 Hz), 1.79 (m, 4H), 3.10 (t, 2H, *J* = 7.6 Hz), 3.63 (t, 2H, *J* = 8.0 Hz), 7.08 (t, 1H, *J* = 9.6 Hz), 7.11 (t, 1H, *J* = 9.6 Hz), 7.32 (d, 1H, *J* = 4.0 Hz), 7.49 (m, 4H), 7.57 (t, 1H, *J* = 9.8 Hz), 7.79 (d, 1H, *J* = 3.6 Hz), 8.04 (d, 2H, *J* = 8.4 Hz), 8.27 (d, 1H, *J* = 9.6 Hz), 8.32 (d, 2H, *J* = 9.2 Hz), 8.34 (d, 1H, *J* = 9.6 Hz), 8.41 (s, 1H); MS (MALDI-TOF) m/z 384 (M+).

**Figure S1**

Optimized structures of bridge azulene-anthracene's in vacuo were computed by density functional method (B3LYP/6-31G*) starting from five most stable conformations determined by semi-empirical molecular orbital method (PM5). Black, red and blue spheres are carbon, oxygen and hydrogen atoms, respectively. Tortional angles around C-C single bond along the bridge structure are displayed. (a) 9-(6-Azulen-1-yl-ethyl)-anthracene. (b) 9-(6-Azulen-1-yl-propyl)-anthracene. (c) 9-(6-Azulen-1-yl-hexyl)-anthracene. (d) 9-(Azulen-1-ylmethoxymethyl)-anthracene. (e) 9-[2-(Azulen-1-ylmethoxy)-ethoxymethyl]-anthracene.

The optimized electronic energy for each state is as follows (the unit is eV):

(a) 1. -27287.424, 2. -27287.377, 3. -27287.390, 4. -27287.376.
(b) 1. -28357.202, 2. -28357.198, 3. -28357.199, 4. -28357.199, 5. -28357.164.
(c) 1. -31566.545, 2. -31566.545, 3. -31566.528, 4. -31566.528, 5. -31566.510.
(d) 1. -29333.810, 2. -29333.761, 3. -29333.808, 4. -29333.808, 5. -29333.810.
(e) 1. -33519.681, 2. -33519.575, 3. -33519.588, 4. -33519.602, 5. -33519.636.

(a)

1 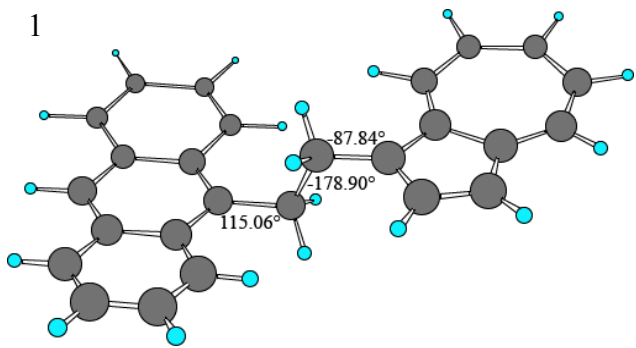 3 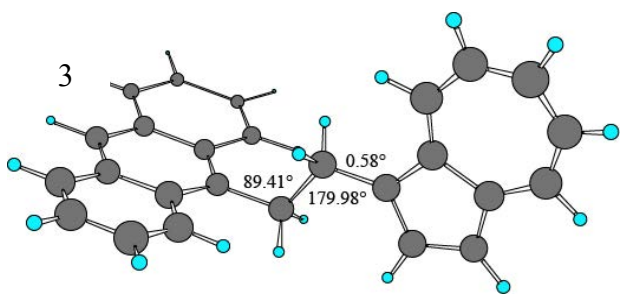

2 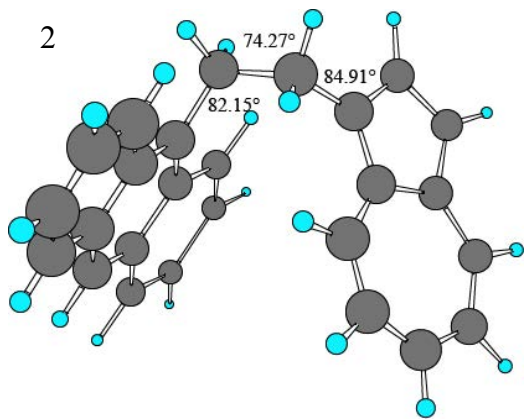 4 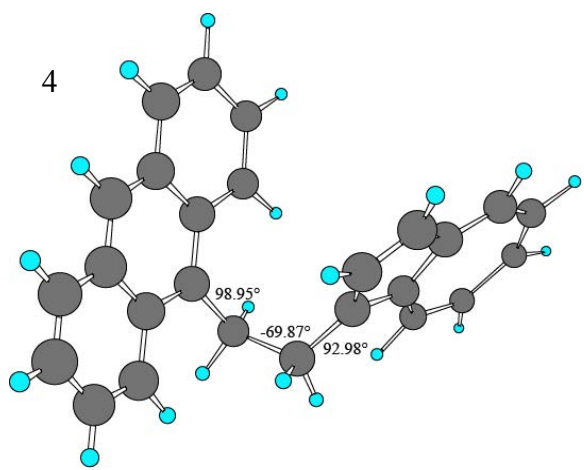

(b)

1
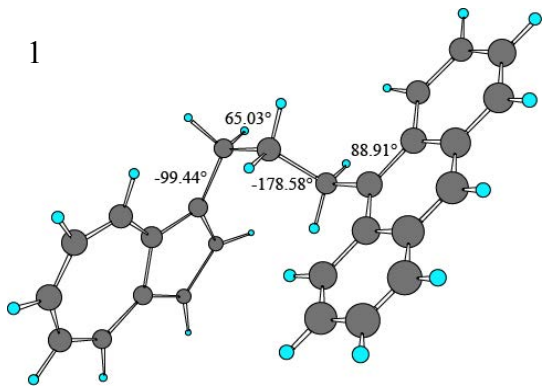

4
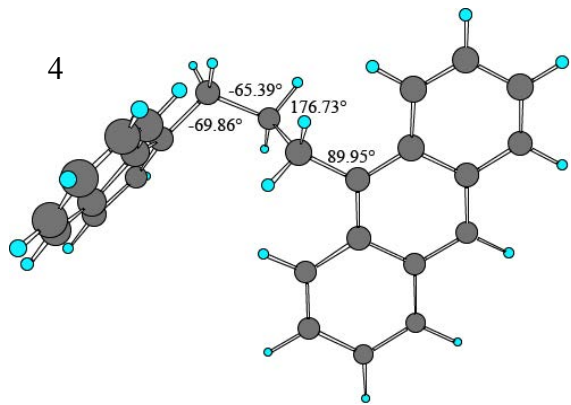

2
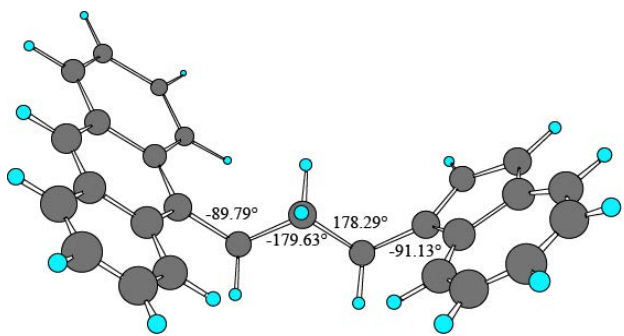

5
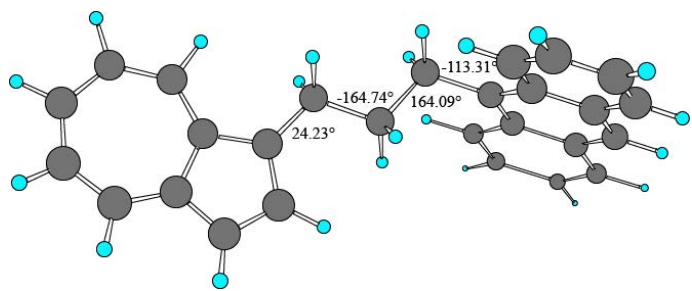

3
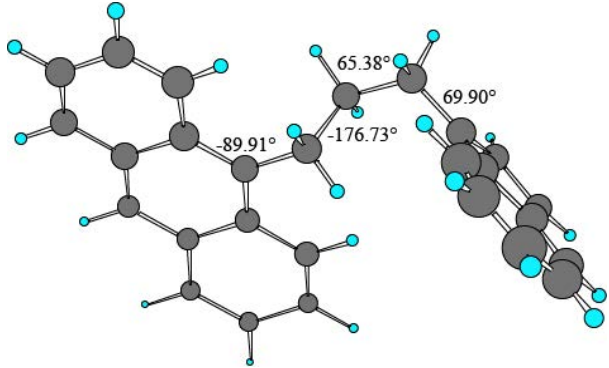

(c)

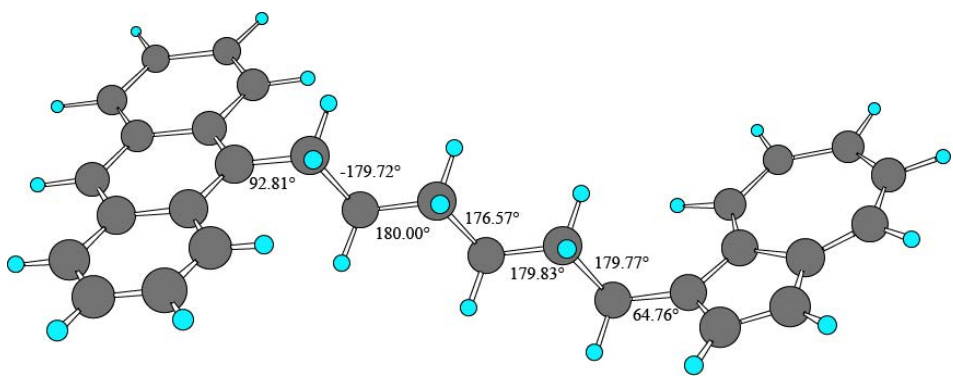

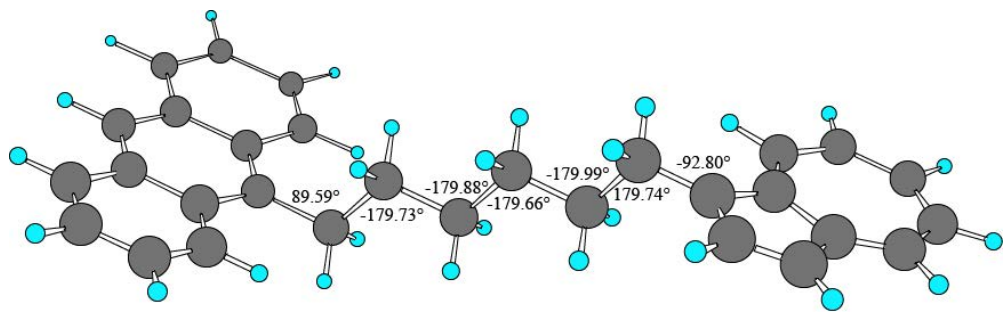

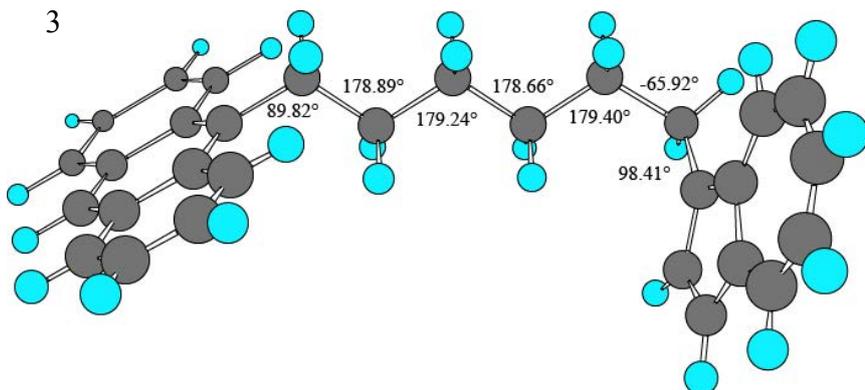

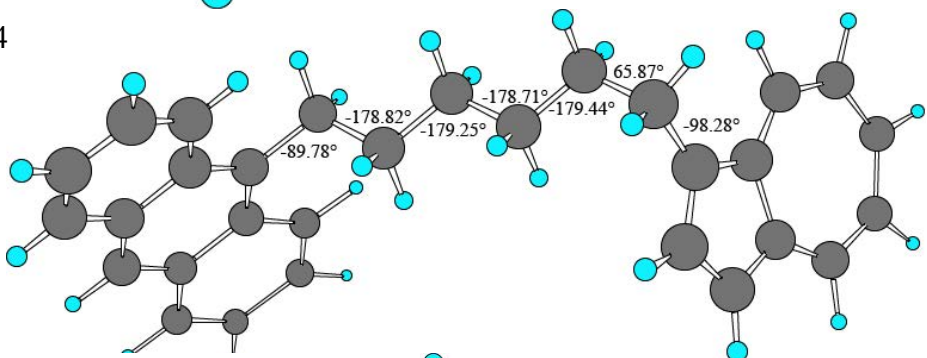

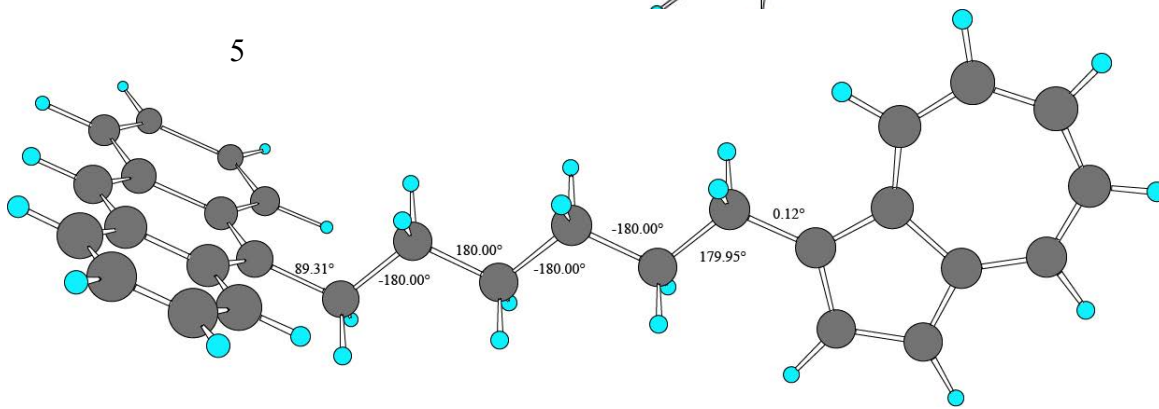

(d)

1
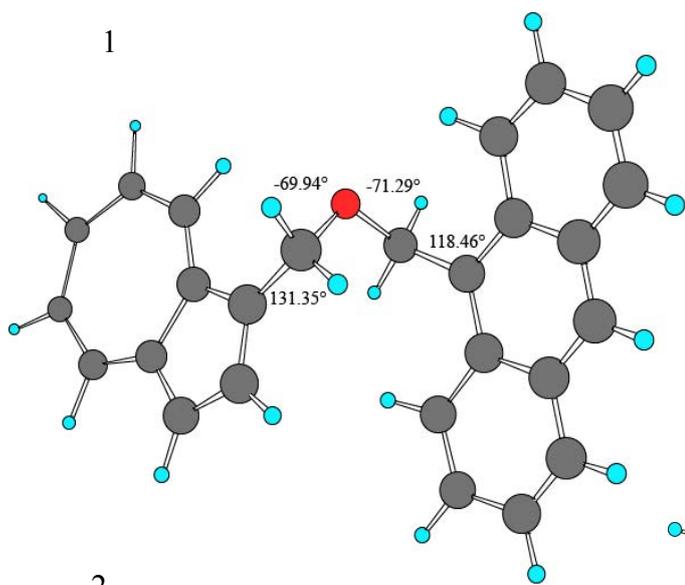

2
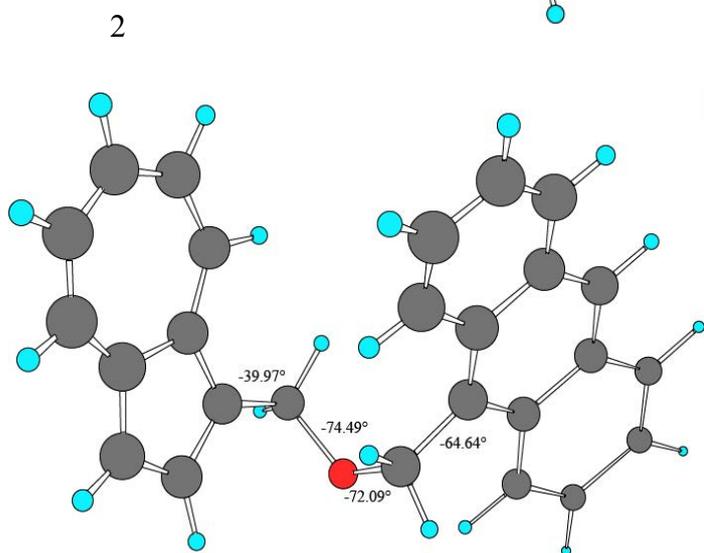

3
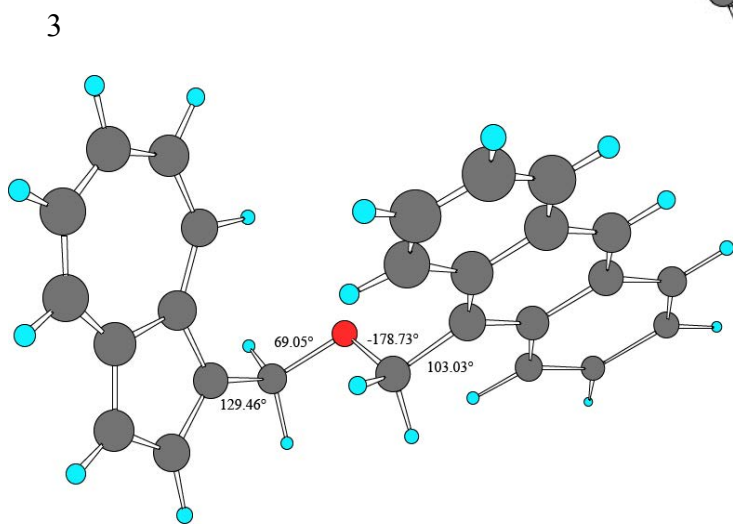

4
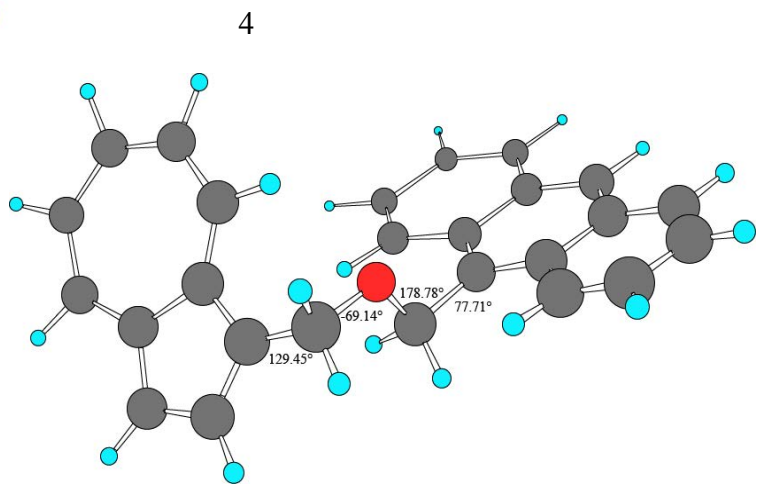

5
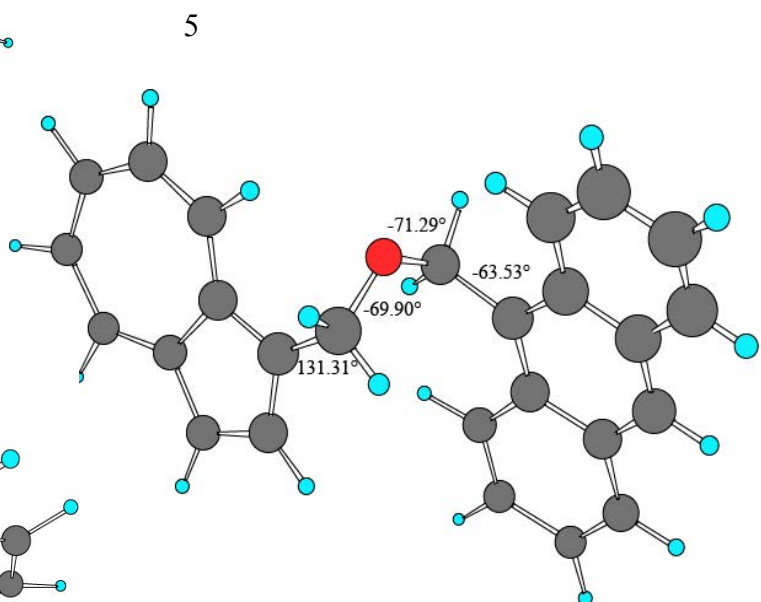

(e)

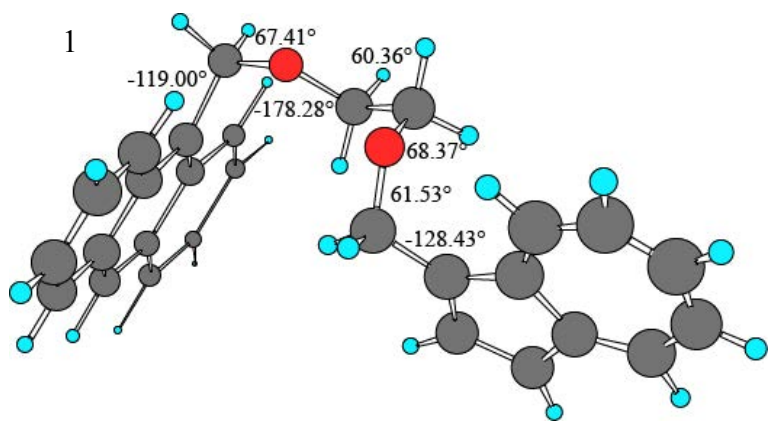

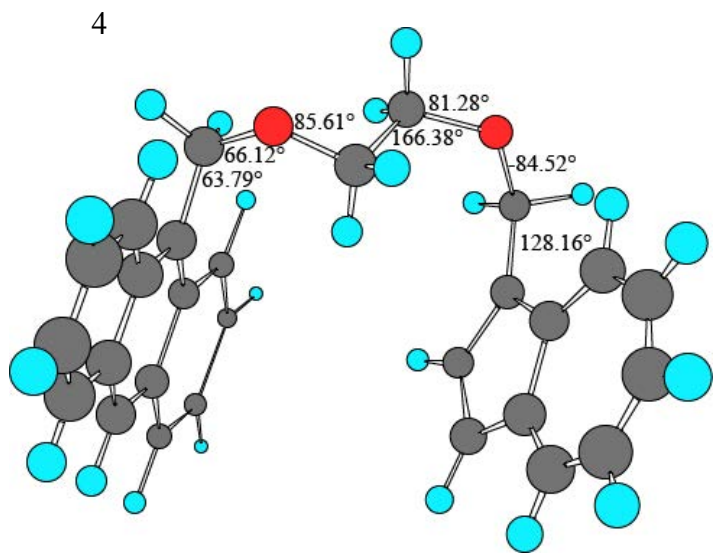

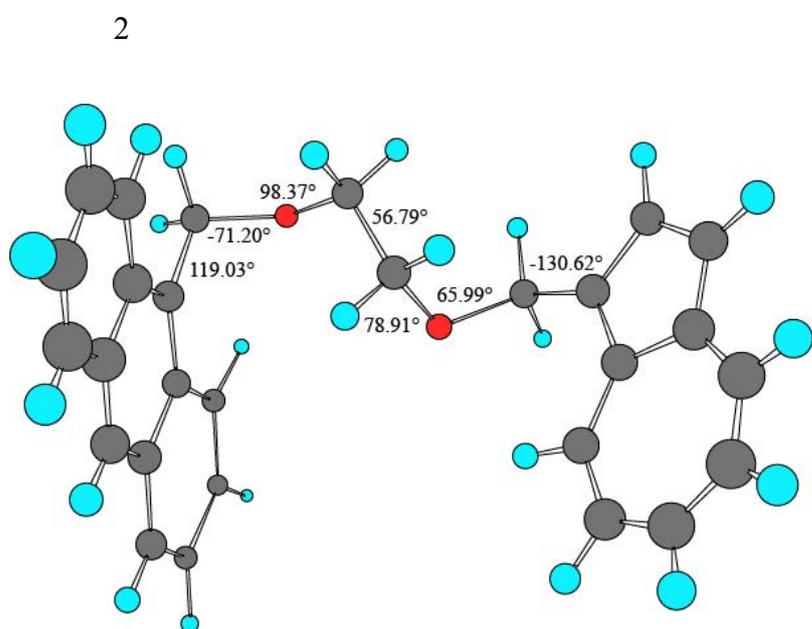

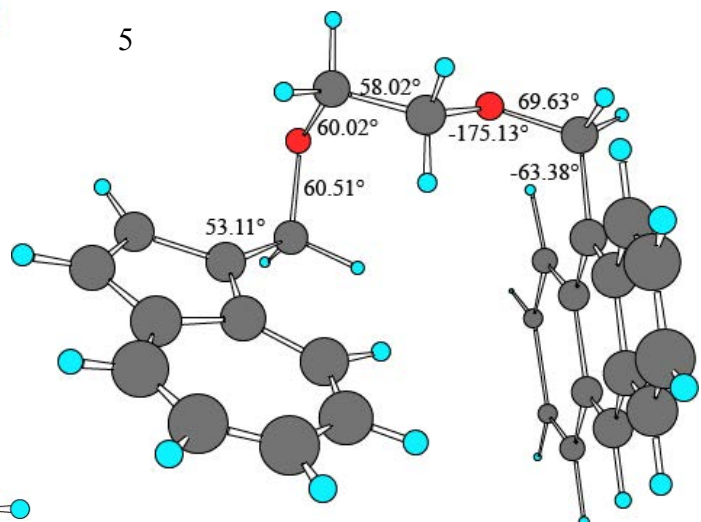

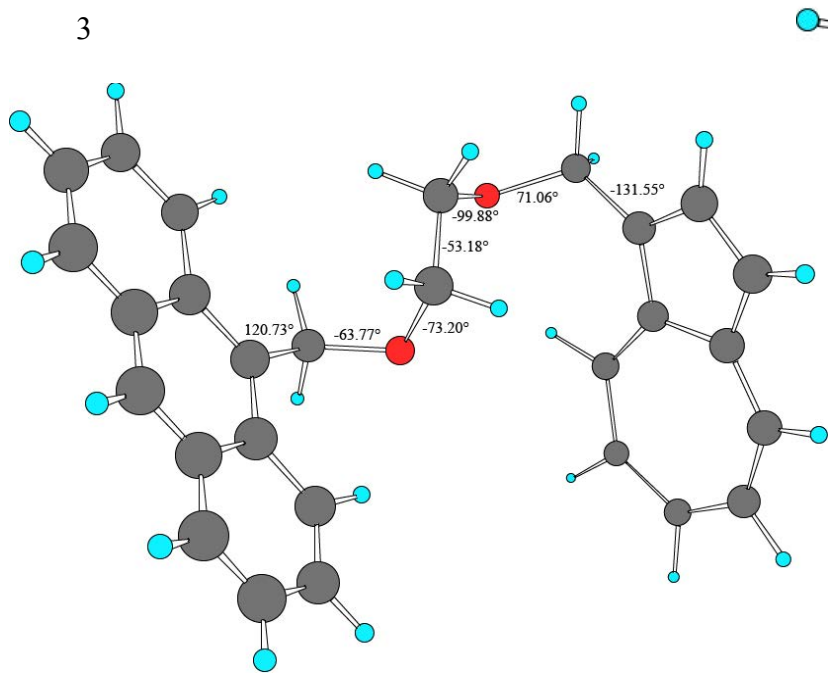

**Table S1**

Vibrational normal modes of AzEtAn.[a]

| Frequency [cm⁻¹] | | | IR Intensity [km/mol] | | | | |
|---|---|---|---|---|---|---|---|
| 13.2 | ± | 4.3 | 4 | (± | 1 | )× 10 | ⁻² |
| 23.0 | ± | 2.9 | 5 | (± | 1 | )× 10 | ⁻² |
| 37.9 | ± | 0.8 | 1 | (± | 1 | )× 10 | ⁻¹ |
| 53.9 | ± | 14.0 | 4 | (± | 8 | )× 10 | ⁻² |
| 77.2 | ± | 4.7 | 2 | (± | 2 | )× 10 | ⁻¹ |
| 92.0 | ± | 2.5 | 5 | (± | 3 | )× 10 | ⁻¹ |
| 116.8 | ± | 3.4 | 5 | (± | 8 | )× 10 | ⁻² |
| 135.2 | ± | 4.2 | 1 | (± | 2 | )× 10 | ⁻¹ |
| 175.0 | ± | 1.7 | 7 | (± | 5 | )× 10 | ⁻¹ |
| 188.8 | ± | 12.3 | 5 | (± | 2 | )× 10 | ⁻¹ |
| 239.1 | ± | 0.5 | 5 | (± | 6 | )× 10 | ⁻² |
| 248.8 | ± | 2.7 | 70 | (± | 8 | )× 10 | ⁻² |
| 265.4 | ± | 8.3 | 10 | (± | 5 | )× 10 | ⁻¹ |
| 306.9 | ± | 1.5 | 3 | (± | 1 | )× 10 | ⁻¹ |
| 328.5 | ± | 1.4 | 58 | (± | 6 | )× 10 | ⁻¹ |
| 349.5 | ± | 12.7 | 64 | (± | 10 | )× 10 | ⁻² |
| 378.5 | ± | 12.8 | 4 | (± | 5 | )× 10 | ⁻¹ |
| 398.3 | ± | 0.2 | 7 | (± | 2 | )× 10 | ⁻¹ |
| 399.7 | ± | 0.7 | 4 | (± | 2 | )× 10 | ⁻¹ |
| 418.1 | ± | 1.9 | 2 | (± | 1 | )× 10 | ⁰ |
| 424.4 | ± | 1.4 | 19 | (± | 6 | )× 10 | ⁻¹ |
| 431.9 | ± | 1.1 | 14 | (± | 9 | )× 10 | ⁻¹ |
| 457.8 | ± | 7.8 | 5 | (± | 3 | )× 10 | ⁰ |
| 488.1 | ± | 0.6 | 2 | (± | 2 | )× 10 | ⁻¹ |
| 496.4 | ± | 2.2 | 6 | (± | 4 | )× 10 | ⁰ |
| 508.5 | ± | 0.1 | 2 | (± | 2 | )× 10 | ⁻² |
| 538.6 | ± | 16.3 | 29 | (± | 2 | )× 10 | ⁻¹ |
| 561.7 | ± | 1.2 | 6 | (± | 1 | )× 10 | ⁻¹ |
| 574.9 | ± | 1.6 | 21 | (± | 6 | )× 10 | ⁻¹ |
| 594.2 | ± | 0.8 | 47 | (± | 3 | )× 10 | ⁻¹ |
| 616.7 | ± | 1.3 | 41 | (± | 6 | )× 10 | ⁻¹ |
| 619.1 | ± | 0.8 | 29 | (± | 6 | )× 10 | ⁻¹ |
| 641.1 | ± | 3.0 | 17 | (± | 6 | )× 10 | ⁻¹ |
| 654.9 | ± | 0.1 | 11 | (± | 1 | )× 10 | ⁻¹ |
| 675.1 | ± | 13.2 | 8 | (± | 5 | )× 10 | ⁻¹ |
| 693.0 | ± | 3.2 | 11 | (± | 7 | )× 10 | ⁻¹ |
| 715.4 | ± | 14.9 | 8 | (± | 4 | )× 10 | ⁻¹ |
| 736.8 | ± | 0.8 | 4 | (± | 2 | )× 10 | ⁻¹ |
| 748.6 | ± | 0.6 | 58 | (± | 5 | )× 10 | ⁰ |
| 756.1 | ± | 0.5 | 9 | (± | 6 | )× 10 | ⁻¹ |
| 763.0 | ± | 1.8 | 23 | (± | 8 | )× 10 | ⁰ |
| 771.1 | ± | 1.5 | 4 | (± | 1 | )× 10 | ⁰ |
| 773.0 | ± | 0.3 | 1 | (± | 1 | )× 10 | ¹ |
| 786.1 | ± | 5.1 | 2 | (± | 1 | )× 10 | ¹ |
| 796.1 | ± | 5.0 | 15 | (± | 5 | )× 10 | ⁰ |
| 818.5 | ± | 15.4 | 6 | (± | 4 | )× 10 | ⁰ |
| 852.4 | ± | 3.8 | 3 | (± | 5 | )× 10 | ⁰ |
| 857.4 | ± | 0.6 | 13 | (± | 3 | )× 10 | ⁰ |
| 867.1 | ± | 0.5 | 2 | (± | 1 | )× 10 | ⁻¹ |
| 874.2 | ± | 1.1 | 8 | (± | 2 | )× 10 | ⁻¹ |
| 882.8 | ± | 4.5 | 3 | (± | 2 | )× 10 | ⁰ |
| 898.7 | ± | 2.6 | 11 | (± | 7 | )× 10 | ⁰ |
| 904.8 | ± | 7.1 | 12 | (± | 10 | )× 10 | ⁰ |
| 917.7 | ± | 2.2 | 2 | (± | 1 | )× 10 | ⁰ |
| 924.2 | ± | 4.2 | 2 | (± | 2 | )× 10 | ⁰ |
| 950.4 | ± | 9.5 | 2 | (± | 5 | )× 10 | ⁰ |
| 957.5 | ± | 0.5 | 2 | (± | 2 | )× 10 | ⁰ |
| 967.6 | ± | 6.7 | 34 | (± | 7 | )× 10 | ⁻¹ |
| 978.0 | ± | 4.3 | 4 | (± | 2 | )× 10 | ⁰ |
| 984.8 | ± | 3.0 | 2 | (± | 3 | )× 10 | ⁰ |
| 987.2 | ± | 0.5 | 5 | (± | 4 | )× 10 | ⁻¹ |
| 989.4 | ± | 1.1 | 2 | (± | 2 | )× 10 | ⁰ |
| 995.2 | ± | 1.5 | 10 | (± | 8 | )× 10 | ⁻¹ |
| 1,004.5 | ± | 2.1 | 1 | (± | 3 | )× 10 | ⁰ |
| 1,010.2 | ± | 5.6 | 10 | (± | 10 | )× 10 | ⁻¹ |
| 1,035.4 | ± | 1.5 | 5 | (± | 3 | )× 10 | ⁰ |
| 1,047.3 | ± | 3.6 | 9 | (± | 7 | )× 10 | ⁰ |
| 1,052.0 | ± | 1.7 | 2 | (± | 3 | )× 10 | ⁰ |
| 1,060.9 | ± | 2.9 | 33 | (± | 9 | )× 10 | ⁻¹ |
| 1,074.6 | ± | 2.6 | 37 | (± | 10 | )× 10 | ⁻¹ |
| 1,081.9 | ± | 3.1 | 3 | (± | 1 | )× 10 | ⁰ |
| 1,135.6 | ± | 0.1 | 10 | (± | 1 | )× 10 | ⁻² |
| 1,166.2 | ± | 6.0 | 6 | (± | 2 | )× 10 | ⁻¹ |
| 1,189.8 | ± | 0.7 | 6 | (± | 6 | )× 10 | ⁻¹ |
| 1,192.2 | ± | 0.4 | 3 | (± | 1 | )× 10 | ⁰ |
| 1,205.0 | ± | 2.4 | 11 | (± | 8 | )× 10 | ⁻¹ |
| 1,211.6 | ± | 1.0 | 53 | (± | 9 | )× 10 | ⁻² |
| 1,222.4 | ± | 3.1 | 28 | (± | 10 | )× 10 | ⁻¹ |
| 1,247.0 | ± | 4.5 | 14 | (± | 7 | )× 10 | ⁻¹ |
| **1,258.1** | ± | 0.7 | 3 | (± | 4 | )× 10 | -1 |
| **1,264.6** | ± | 1.4 | 7 | (± | 1 | )× 10 | -1 |
| **1,281.3** | ± | 10.3 | 29 | (± | 7 | )× 10 | -1 |
| **1,298.0** | ± | 3.0 | 20 | (± | 2 | )× 10 | -1 |
| **1,313.5** | ± | 0.8 | 2 | (± | 2 | )× 10 | 0 |
| **1,323.9** | ± | 11.6 | 4 | (± | 2 | )× 10 | -1 |
| **1,337.5** | ± | 3.9 | 1 | (± | 2 | )× 10 | 0 |
| **1,346.3** | ± | 5.9 | 9 | (± | 2 | )× 10 | 0 |
| **1,359.9** | ± | 1.7 | 0 | (± | 1 | )× 10 | 1 |
| **1,372.8** | ± | 2.2 | 4 | (± | 5 | )× 10 | 0 |
| **1,379.6** | ± | 4.0 | 12 | (± | 7 | )× 10 | 0 |
| **1,387.7** | ± | 9.9 | 62 | (± | 8 | )× 10 | -1 |
| **1,416.6** | ± | 0.2 | 19 | (± | 1 | )× 10 | -1 |
| **1,430.1** | ± | 0.8 | 93 | (± | 5 | )× 10 | -2 |
| **1,439.7** | ± | 1.0 | 7 | (± | 1 | )× 10 | 1 |
| **1,450.1** | ± | 0.3 | 5 | (± | 3 | )× 10 | 0 |
| **1,469.2** | ± | 1.2 | 7 | (± | 2 | )× 10 | 0 |
| **1,488.9** | ± | 0.6 | 14 | (± | 4 | )× 10 | 0 |
| **1,494.3** | ± | 0.2 | 10 | (± | 4 | )× 10 | -1 |
| **1,497.6** | ± | 0.8 | 43 | (± | 5 | )× 10 | -1 |
| **1,508.3** | ± | 0.3 | 7 | (± | 1 | )× 10 | 0 |
| **1,520.8** | ± | 1.9 | 5 | (± | 1 | )× 10 | 0 |
| **1,529.8** | ± | 1.4 | 20 | (± | 5 | )× 10 | -1 |
| **1,558.6** | ± | 3.2 | 5 | (± | 2 | )× 10 | 0 |
| **1,568.7** | ± | 1.4 | 8 | (± | 2 | )× 10 | 0 |
| **1,579.1** | ± | 0.2 | 30 | (± | 2 | )× 10 | -1 |
| **1,596.0** | ± | 0.3 | 33 | (± | 3 | )× 10 | -1 |
| **1,609.7** | ± | 0.2 | 9 | (± | 6 | )× 10 | -1 |
| **1,633.4** | ± | 0.3 | 13 | (± | 2 | )× 10 | -1 |
| **1,647.2** | ± | 0.9 | 72 | (± | 7 | )× 10 | 0 |
| **1,661.7** | ± | 0.2 | 6 | (± | 1 | )× 10 | 0 |
| **1,677.1** | ± | 0.2 | 1 | (± | 1 | )× 10 | -1 |
| **1,686.4** | ± | 0.3 | 66 | (± | 2 | )× 10 | -1 |
| **3,043.2** | ± | 5.2 | 26 | (± | 7 | )× 10 | 0 |
| **3,075.0** | ± | 6.8 | 2 | (± | 2 | )× 10 | 1 |
| **3,084.7** | ± | 1.1 | 23 | (± | 7 | )× 10 | 0 |
| **3,117.7** | ± | 0.7 | 15 | (± | 5 | )× 10 | 0 |
| **3,150.0** | ± | 0.7 | 6 | (± | 2 | )× 10 | 0 |
| **3,154.7** | ± | 1.5 | 11 | (± | 2 | )× 10 | 0 |
| **3,163.2** | ± | 4.6 | 10 | (± | 4 | )× 10 | 0 |
| **3,173.4** | ± | 0.9 | 34 | (± | 2 | )× 10 | -1 |
| **3,178.2** | ± | 0.5 | 10 | (± | 5 | )× 10 | -2 |
| **3,179.3** | ± | 0.2 | 3 | (± | 2 | )× 10 | 1 |
| **3,180.2** | ± | 1.1 | 2 | (± | 2 | )× 10 | 1 |
| **3,188.4** | ± | 0.4 | 2 | (± | 1 | )× 10 | 1 |
| **3,188.6** | ± | 0.5 | 24 | (± | 6 | )× 10 | 0 |
| **3,189.9** | ± | 0.6 | 20 | (± | 2 | )× 10 | 0 |
| **3,204.8** | ± | 0.9 | 4 | (± | 2 | )× 10 | 1 |
| **3,205.5** | ± | 0.3 | 4 | (± | 1 | )× 10 | 1 |
| **3,207.4** | ± | 7.5 | 19 | (± | 9 | )× 10 | 0 |
| **3,223.5** | ± | 1.0 | 22 | (± | 2 | )× 10 | 0 |
| **3,229.1** | ± | 7.7 | 20 | (± | 5 | )× 10 | 0 |
| **3,238.5** | ± | 2.0 | 13 | (± | 1 | )× 10 | 0 |

a B3LYP/6-31G(d)

b Canonical ensemble average of five most stable configurations. The values following "±" sign is square root of Canonical ensemble variance.

c infrared

**Table S2**

Vibrational normal modes of AzPrAn[a]

| Frequency [cm$^{-1}$] | | | IR Intensity [km/mol] | | | | | |
|---|---|---|---|---|---|---|---|---|
| 10.0 | ± | 2.0 | 46 | (± | 9 | )× | 10 | $^{-3}$ |
| 16.3 | ± | 1.4 | 3 | (± | 3 | )× | 10 | $^{-2}$ |
| 27.2 | ± | 4.8 | 5 | (± | 3 | )× | 10 | $^{-2}$ |
| 50.2 | ± | 4.2 | 19 | (± | 6 | )× | 10 | $^{-3}$ |
| 64.1 | ± | 7.0 | 6 | (± | 5 | )× | 10 | $^{-2}$ |
| 90.1 | ± | 1.1 | 5 | (± | 2 | )× | 10 | $^{-1}$ |
| 103.1 | ± | 3.7 | 3 | (± | 2 | )× | 10 | $^{-1}$ |
| 119.8 | ± | 4.7 | 6 | (± | 9 | )× | 10 | $^{-2}$ |
| 156.6 | ± | 22.3 | 5 | (± | 3 | )× | 10 | $^{-1}$ |
| 172.6 | ± | 16.8 | 7 | (± | 3 | )× | 10 | $^{-1}$ |
| 191.8 | ± | 10.5 | 3 | (± | 3 | )× | 10 | $^{-1}$ |
| 224.4 | ± | 4.6 | 15 | (± | 8 | )× | 10 | $^{-1}$ |
| 239.5 | ± | 0.2 | 9 | (± | 3 | )× | 10 | $^{-3}$ |
| 259.4 | ± | 6.6 | 4 | (± | 2 | )× | 10 | $^{-1}$ |
| 278.3 | ± | 9.8 | 6 | (± | 7 | )× | 10 | $^{-1}$ |
| 317.8 | ± | 14.4 | 2 | (± | 1 | )× | 10 | $^{0}$ |
| 327.2 | ± | 1.0 | 4 | (± | 1 | )× | 10 | $^{0}$ |
| 368.8 | ± | 12.2 | 6 | (± | 7 | )× | 10 | $^{-1}$ |
| 382.2 | ± | 7.0 | 7 | (± | 5 | )× | 10 | $^{-1}$ |
| 398.8 | ± | 0.7 | 4 | (± | 3 | )× | 10 | $^{-1}$ |
| 400.6 | ± | 1.1 | 7 | (± | 4 | )× | 10 | $^{-1}$ |
| 422.4 | ± | 2.0 | 3 | (± | 1 | )× | 10 | $^{0}$ |
| 425.4 | ± | 0.8 | 1 | (± | 1 | )× | 10 | $^{0}$ |
| 430.9 | ± | 5.8 | 2 | (± | 2 | )× | 10 | $^{0}$ |
| 463.0 | ± | 6.1 | 5 | (± | 3 | )× | 10 | $^{0}$ |
| 489.5 | ± | 0.3 | 7 | (± | 3 | )× | 10 | $^{-2}$ |
| 497.3 | ± | 5.8 | 234 | (± | 9 | )× | 10 | $^{-2}$ |
| 508.6 | ± | 0.2 | 6 | (± | 2 | )× | 10 | $^{-3}$ |
| 552.2 | ± | 0.4 | 60 | (± | 4 | )× | 10 | $^{-1}$ |
| 562.8 | ± | 0.4 | 38 | (± | 8 | )× | 10 | $^{-2}$ |
| 585.8 | ± | 2.8 | 5 | (± | 2 | )× | 10 | $^{-1}$ |
| 591.6 | ± | 3.7 | 46 | (± | 7 | )× | 10 | $^{-1}$ |
| 617.2 | ± | 0.5 | 3 | (± | 1 | )× | 10 | $^{0}$ |
| 619.4 | ± | 3.2 | 3 | (± | 2 | )× | 10 | $^{0}$ |
| 650.9 | ± | 1.6 | 20 | (± | 6 | )× | 10 | $^{-1}$ |
| 657.4 | ± | 2.6 | 20 | (± | 5 | )× | 10 | $^{-1}$ |
| 670.1 | ± | 11.1 | 10 | (± | 9 | )× | 10 | $^{-1}$ |
| 693.1 | ± | 6.4 | 7 | (± | 6 | )× | 10 | $^{-1}$ |
| 716.3 | ± | 7.8 | 16 | (± | 10 | )× | 10 | $^{-1}$ |
| 734.6 | ± | 1.1 | 4 | (± | 5 | )× | 10 | $^{-1}$ |
| 747.8 | ± | 2.7 | 3 | (± | 2 | )× | 10 | $^{1}$ |
| 754.0 | ± | 3.8 | 2 | (± | 3 | )× | 10 | $^{1}$ |
| 757.9 | ± | 1.7 | 1 | (± | 1 | )× | 10 | $^{1}$ |
| 767.8 | ± | 4.0 | 2 | (± | 1 | )× | 10 | $^{1}$ |
| 773.1 | ± | 0.8 | 4 | (± | 8 | )× | 10 | $^{-1}$ |
| 783.2 | ± | 5.6 | 2 | (± | 1 | )× | 10 | $^{1}$ |
| 792.8 | ± | 2.5 | 23 | (± | 9 | )× | 10 | $^{0}$ |
| 804.5 | ± | 0.6 | 75 | (± | 4 | )× | 10 | $^{-1}$ |
| 836.7 | ± | 12.5 | 8 | (± | 1 | )× | 10 | $^{-1}$ |
| 856.2 | ± | 1.3 | 8 | (± | 4 | )× | 10 | $^{0}$ |
| 867.3 | ± | 1.3 | 3 | (± | 3 | )× | 10 | $^{0}$ |
| 870.5 | ± | 1.8 | 3 | (± | 2 | )× | 10 | $^{0}$ |
| 875.7 | ± | 1.3 | 17 | (± | 10 | )× | 10 | $^{-1}$ |
| 883.5 | ± | 2.4 | 3 | (± | 2 | )× | 10 | $^{0}$ |
| 894.3 | ± | 3.0 | 10 | (± | 3 | )× | 10 | $^{0}$ |
| 900.4 | ± | 0.2 | 18 | (± | 2 | )× | 10 | $^{0}$ |
| 921.2 | ± | 0.8 | 2 | (± | 1 | )× | 10 | $^{0}$ |
| 927.0 | ± | 4.5 | 2 | (± | 2 | )× | 10 | $^{0}$ |
| 956.1 | ± | 0.5 | 6 | (± | 4 | )× | 10 | $^{-1}$ |
| 958.7 | ± | 2.1 | 22 | (± | 5 | )× | 10 | $^{-1}$ |
| 967.1 | ± | 4.9 | 3 | (± | 1 | )× | 10 | $^{0}$ |
| 976.3 | ± | 3.2 | 3 | (± | 1 | )× | 10 | $^{0}$ |
| 982.0 | ± | 2.7 | 4 | (± | 2 | )× | 10 | $^{0}$ |
| 986.7 | ± | 0.8 | 14 | (± | 8 | )× | 10 | $^{-2}$ |
| 990.4 | ± | 1.4 | 18 | (± | 6 | )× | 10 | $^{-2}$ |
| 994.5 | ± | 1.5 | 6 | (± | 10 | )× | 10 | $^{-1}$ |
| 1,005.3 | ± | 0.5 | 1 | (± | 1 | )× | 10 | $^{0}$ |
| 1,021.1 | ± | 8.3 | 4 | (± | 2 | )× | 10 | $^{0}$ |
| 1,039.2 | ± | 3.2 | 7 | (± | 2 | )× | 10 | $^{0}$ |
| 1,044.1 | ± | 2.8 | 5 | (± | 2 | )× | 10 | $^{0}$ |
| 1,050.4 | ± | 1.7 | 2 | (± | 1 | )× | 10 | $^{0}$ |
| 1,059.4 | ± | 4.5 | 1 | (± | 2 | )× | 10 | $^{0}$ |
| 1,066.8 | ± | 4.8 | 4 | (± | 1 | )× | 10 | $^{0}$ |
| 1,077.9 | ± | 2.2 | 43 | (± | 4 | )× | 10 | $^{-1}$ |
| 1,097.2 | ± | 4.1 | 4 | (± | 3 | )× | 10 | $^{0}$ |
| 1,135.8 | ± | 0.2 | 11 | (± | 4 | )× | 10 | $^{-2}$ |
| 1,161.3 | ± | 4.1 | 6 | (± | 3 | )× | 10 | $^{-1}$ |
| 1,189.0 | ± | 1.2 | 2 | (± | 2 | )× | 10 | $^{-1}$ |
| 1,191.9 | ± | 0.1 | 26 | (± | 2 | )× | 10 | $^{-1}$ |

| | | | | | | | |
|---|---|---|---|---|---|---|---|
| **1,203.1** | ± | 2.0 | 15 | (± | 3 | )× 10 | -1 |
| **1,211.9** | ± | 0.2 | 29 | (± | 9 | )× 10 | -2 |
| **1,220.4** | ± | 3.5 | 15 | (± | 5 | )× 10 | -1 |
| **1,235.9** | ± | 6.1 | 5 | (± | 2 | )× 10 | 0 |
| **1,256.6** | ± | 0.4 | 2 | (± | 2 | )× 10 | 0 |
| **1,261.7** | ± | 2.3 | 2 | (± | 1 | )× 10 | 0 |
| **1,268.2** | ± | 4.2 | 2 | (± | 2 | )× 10 | 0 |
| **1,294.6** | ± | 0.6 | 12 | (± | 7 | )× 10 | -1 |
| **1,309.2** | ± | 5.1 | 4 | (± | 1 | )× 10 | 0 |
| **1,317.2** | ± | 1.8 | 9 | (± | 7 | )× 10 | -1 |
| **1,331.1** | ± | 0.5 | 8 | (± | 3 | )× 10 | -1 |
| **1,336.3** | ± | 1.7 | 3 | (± | 3 | )× 10 | -1 |
| **1,341.9** | ± | 1.1 | 8 | (± | 3 | )× 10 | 0 |
| **1,355.3** | ± | 3.7 | 7 | (± | 7 | )× 10 | 0 |
| **1,373.3** | ± | 4.7 | 9 | (± | 2 | )× 10 | 0 |
| **1,377.1** | ± | 1.2 | 5 | (± | 3 | )× 10 | 0 |
| **1,383.1** | ± | 2.2 | 7 | (± | 6 | )× 10 | 0 |
| **1,400.2** | ± | 3.1 | 14 | (± | 9 | )× 10 | 0 |
| **1,417.2** | ± | 0.2 | 23 | (± | 1 | )× 10 | -1 |
| **1,430.4** | ± | 0.2 | 12 | (± | 2 | )× 10 | -1 |
| **1,439.1** | ± | 1.9 | 77 | (± | 4 | )× 10 | 0 |
| **1,450.7** | ± | 0.1 | 56 | (± | 6 | )× 10 | -1 |
| **1,469.0** | ± | 0.4 | 8 | (± | 5 | )× 10 | 0 |
| **1,487.9** | ± | 2.3 | 14 | (± | 4 | )× 10 | 0 |
| **1,494.8** | ± | 0.1 | 14 | (± | 2 | )× 10 | -1 |
| **1,498.3** | ± | 0.4 | 40 | (± | 4 | )× 10 | -1 |
| **1,508.3** | ± | 0.5 | 8 | (± | 1 | )× 10 | 0 |
| **1,515.3** | ± | 0.9 | 7 | (± | 4 | )× 10 | 0 |
| **1,522.7** | ± | 1.7 | 2 | (± | 1 | )× 10 | 0 |
| **1,533.2** | ± | 0.8 | 13 | (± | 5 | )× 10 | -1 |
| **1,561.8** | ± | 2.0 | 52 | (± | 5 | )× 10 | -1 |
| **1,568.8** | ± | 1.8 | 8 | (± | 3 | )× 10 | 0 |
| **1,579.7** | ± | 0.2 | 31 | (± | 1 | )× 10 | -1 |
| **1,595.9** | ± | 0.4 | 31 | (± | 4 | )× 10 | -1 |
| **1,610.4** | ± | 0.2 | 14 | (± | 3 | )× 10 | -1 |
| **1,633.5** | ± | 0.1 | 12 | (± | 1 | )× 10 | -1 |
| **1,647.1** | ± | 0.3 | 75 | (± | 4 | )× 10 | 0 |
| **1,661.8** | ± | 0.4 | 7 | (± | 1 | )× 10 | 0 |
| **1,677.4** | ± | 0.2 | 6 | (± | 3 | )× 10 | -2 |
| **1,686.5** | ± | 0.2 | 69 | (± | 3 | )× 10 | -1 |
| **3,024.0** | ± | 13.1 | 30 | (± | 3 | )× 10 | 0 |
| **3,042.2** | ± | 12.2 | 21 | (± | 9 | )× 10 | 0 |
| **3,060.4** | ± | 8.9 | 2 | (± | 1 | )× 10 | 1 |
| **3,068.0** | ± | 0.9 | 2 | (± | 2 | )× 10 | 1 |
| **3,082.7** | ± | 2.9 | 2 | (± | 1 | )× 10 | 1 |
| **3,107.6** | ± | 4.8 | 27 | (± | 4 | )× 10 | 0 |
| **3,150.2** | ± | 0.3 | 63 | (± | 7 | )× 10 | -1 |
| **3,155.3** | ± | 1.1 | 8 | (± | 2 | )× 10 | 0 |
| **3,162.5** | ± | 1.4 | 114 | (± | 6 | )× 10 | -1 |
| **3,172.8** | ± | 0.5 | 35 | (± | 1 | )× 10 | -1 |
| **3,177.7** | ± | 0.5 | 9 | (± | 4 | )× 10 | -2 |
| **3,179.0** | ± | 0.2 | 2 | (± | 2 | )× 10 | 1 |
| **3,180.1** | ± | 0.3 | 3 | (± | 2 | )× 10 | 1 |
| **3,188.1** | ± | 0.2 | 2 | (± | 1 | )× 10 | 1 |
| **3,188.9** | ± | 0.2 | 3 | (± | 2 | )× 10 | 1 |
| **3,189.2** | ± | 0.2 | 21 | (± | 4 | )× 10 | 0 |
| **3,204.7** | ± | 0.5 | 51 | (± | 3 | )× 10 | 0 |
| **3,205.1** | ± | 0.5 | 32 | (± | 3 | )× 10 | 0 |
| **3,211.1** | ± | 6.1 | 17 | (± | 3 | )× 10 | 0 |
| **3,222.3** | ± | 1.3 | 25 | (± | 2 | )× 10 | 0 |
| **3,224.8** | ± | 0.8 | 14 | (± | 5 | )× 10 | 0 |
| **3,239.4** | ± | 1.5 | 13 | (± | 1 | )× 10 | 0 |

a B3LYP/6-31G(d)
b Canonical ensemble average of five most stable configurations. The values following "±" sign is square root of Canonical ensemble variance.
c infrared

**Table S3**

Vibrational normal modes of AzHeAn[a]

| Frequency [cm$^{-1}$] | | | IR Intensity [km/mol] | | | | | |
|---|---|---|---|---|---|---|---|---|
| 9.5 | ± | 2.0 | 9 | (± | 7 | )× | 10 | $^{-3}$ |
| 12.3 | ± | 2.0 | 3 | (± | 1 | )× | 10 | $^{-2}$ |
| 23.0 | ± | 2.1 | 1 | (± | 2 | )× | 10 | $^{-2}$ |
| 26.6 | ± | 0.8 | 15 | (± | 7 | )× | 10 | $^{-3}$ |
| 47.0 | ± | 4.2 | 8 | (± | 7 | )× | 10 | $^{-3}$ |
| 72.3 | ± | 5.4 | 4 | (± | 1 | )× | 10 | $^{-2}$ |
| 79.3 | ± | 1.5 | 2 | (± | 1 | )× | 10 | $^{-1}$ |
| 94.9 | ± | 4.1 | 14 | (± | 9 | )× | 10 | $^{-2}$ |
| 103.8 | ± | 4.9 | 3 | (± | 2 | )× | 10 | $^{-1}$ |
| 115.8 | ± | 0.2 | 1 | (± | 1 | )× | 10 | $^{-1}$ |
| 123.0 | ± | 5.2 | 3 | (± | 3 | )× | 10 | $^{-1}$ |
| 147.0 | ± | 7.8 | 1 | (± | 4 | )× | 10 | $^{-1}$ |
| 156.8 | ± | 8.2 | 8 | (± | 10 | )× | 10 | $^{-2}$ |
| 171.3 | ± | 3.6 | 4 | (± | 2 | )× | 10 | $^{-1}$ |
| 177.2 | ± | 2.3 | 5 | (± | 2 | )× | 10 | $^{-1}$ |
| 217.2 | ± | 5.9 | 4 | (± | 4 | )× | 10 | $^{-1}$ |
| 237.4 | ± | 3.7 | 1 | (± | 3 | )× | 10 | $^{-2}$ |
| 251.4 | ± | 7.2 | 5 | (± | 2 | )× | 10 | $^{-1}$ |
| 266.4 | ± | 7.1 | 9 | (± | 5 | )× | 10 | $^{-1}$ |
| 299.6 | ± | 13.8 | 6 | (± | 4 | )× | 10 | $^{-1}$ |
| 327.0 | ± | 3.3 | 5 | (± | 2 | )× | 10 | $^{0}$ |
| 336.6 | ± | 5.1 | 1 | (± | 1 | )× | 10 | $^{0}$ |
| 376.9 | ± | 1.3 | 12 | (± | 7 | )× | 10 | $^{-3}$ |
| 383.7 | ± | 2.5 | 7 | (± | 5 | )× | 10 | $^{-1}$ |
| 398.1 | ± | 3.0 | 4 | (± | 1 | )× | 10 | $^{-1}$ |
| 401.7 | ± | 1.2 | 14 | (± | 6 | )× | 10 | $^{-2}$ |
| 414.8 | ± | 2.9 | 17 | (± | 8 | )× | 10 | $^{-1}$ |
| 425.9 | ± | 0.4 | 37 | (± | 10 | )× | 10 | $^{-3}$ |
| 433.6 | ± | 4.4 | 3 | (± | 2 | )× | 10 | $^{0}$ |
| 441.7 | ± | 1.1 | 3 | (± | 2 | )× | 10 | $^{0}$ |
| 474.3 | ± | 7.8 | 10 | (± | 8 | )× | 10 | $^{-1}$ |
| 489.2 | ± | 2.3 | 3 | (± | 8 | )× | 10 | $^{-1}$ |
| 507.6 | ± | 2.0 | 0 | (± | 1 | )× | 10 | $^{0}$ |
| 511.3 | ± | 1.3 | 2 | (± | 1 | )× | 10 | $^{0}$ |
| 557.3 | ± | 0.2 | 50 | (± | 3 | )× | 10 | $^{-1}$ |
| 562.6 | ± | 0.4 | 407 | (± | 2 | )× | 10 | $^{-3}$ |
| 588.4 | ± | 1.9 | 2 | (± | 1 | )× | 10 | $^{0}$ |
| 593.7 | ± | 2.3 | 5 | (± | 2 | )× | 10 | $^{0}$ |
| 617.4 | ± | 0.3 | 4 | (± | 2 | )× | 10 | $^{0}$ |
| 624.5 | ± | 6.3 | 1 | (± | 2 | )× | 10 | $^{0}$ |
| 651.3 | ± | 1.2 | 2 | (± | 1 | )× | 10 | $^{0}$ |
| 657.1 | ± | 4.2 | 25 | (± | 2 | )× | 10 | $^{-1}$ |
| 673.6 | ± | 13.6 | 1 | (± | 1 | )× | 10 | $^{0}$ |
| 698.3 | ± | 8.3 | 10 | (± | 6 | )× | 10 | $^{-1}$ |
| 714.0 | ± | 4.3 | 20 | (± | 10 | )× | 10 | $^{-1}$ |
| 735.7 | ± | 1.2 | 4 | (± | 6 | )× | 10 | $^{-1}$ |
| 739.2 | ± | 0.3 | 29 | (± | 7 | )× | 10 | $^{-1}$ |
| 748.2 | ± | 1.0 | 2 | (± | 3 | )× | 10 | $^{1}$ |
| 751.4 | ± | 4.1 | 4 | (± | 3 | )× | 10 | $^{1}$ |
| 756.4 | ± | 1.0 | 3 | (± | 5 | )× | 10 | $^{0}$ |
| 763.8 | ± | 1.9 | 20 | (± | 9 | )× | 10 | $^{0}$ |
| 772.1 | ± | 0.6 | 4 | (± | 7 | )× | 10 | $^{-1}$ |
| 776.0 | ± | 2.6 | 13 | (± | 7 | )× | 10 | $^{0}$ |
| 790.4 | ± | 2.9 | 2 | (± | 1 | )× | 10 | $^{1}$ |
| 797.1 | ± | 7.1 | 10 | (± | 2 | )× | 10 | $^{0}$ |
| 810.3 | ± | 7.8 | 6 | (± | 2 | )× | 10 | $^{0}$ |
| 845.7 | ± | 8.3 | 1 | (± | 1 | )× | 10 | $^{0}$ |
| 857.6 | ± | 0.5 | 12 | (± | 1 | )× | 10 | $^{0}$ |
| 866.7 | ± | 0.5 | 7 | (± | 10 | )× | 10 | $^{-1}$ |
| 871.6 | ± | 2.5 | 253 | (± | 5 | )× | 10 | $^{-2}$ |
| 877.2 | ± | 0.8 | 42 | (± | 4 | )× | 10 | $^{-1}$ |
| 885.2 | ± | 6.4 | 17 | (± | 7 | )× | 10 | $^{-1}$ |
| 897.3 | ± | 2.4 | 4 | (± | 2 | )× | 10 | $^{0}$ |
| 899.9 | ± | 0.6 | 203 | (± | 8 | )× | 10 | $^{-1}$ |
| 920.3 | ± | 0.7 | 14 | (± | 10 | )× | 10 | $^{-1}$ |
| 927.8 | ± | 3.2 | 2 | (± | 2 | )× | 10 | $^{0}$ |
| 955.7 | ± | 0.3 | 4 | (± | 5 | )× | 10 | $^{-1}$ |
| 957.5 | ± | 0.9 | 2 | (± | 1 | )× | 10 | $^{0}$ |
| 966.8 | ± | 7.4 | 32 | (± | 9 | )× | 10 | $^{-1}$ |
| 977.0 | ± | 3.8 | 3 | (± | 1 | )× | 10 | $^{0}$ |
| 982.1 | ± | 1.4 | 3 | (± | 2 | )× | 10 | $^{0}$ |
| 986.3 | ± | 0.1 | 3 | (± | 1 | )× | 10 | $^{-3}$ |
| 987.5 | ± | 0.1 | 32 | (± | 10 | )× | 10 | $^{-2}$ |
| 993.3 | ± | 0.2 | 4 | (± | 3 | )× | 10 | $^{-1}$ |
| 998.3 | ± | 3.5 | 12 | (± | 9 | )× | 10 | $^{-1}$ |
| 1,003.4 | ± | 0.9 | 3 | (± | 1 | )× | 10 | $^{-1}$ |
| 1,006.6 | ± | 1.0 | 1 | (± | 1 | )× | 10 | $^{0}$ |
| 1,022.0 | ± | 5.1 | 21 | (± | 8 | )× | 10 | $^{-1}$ |
| 1,040.3 | ± | 1.4 | 11 | (± | 3 | )× | 10 | $^{0}$ |

| | | | | | | | | | | | | | |
|---|---|---|---|---|---|---|---|---|---|---|---|---|---|
| **1,048.6** | ± | 0.9 | 4 | (± | 1 | )× | 10 | $^{0}$ | **1,494.7** | ± | 0.3 | 109 | (± | 3 | )× | 10 | $^{-2}$ |
| **1,050.9** | ± | 0.3 | 3 | (± | 3 | )× | 10 | $^{0}$ | **1,498.0** | ± | 0.1 | 447 | (± | 5 | )× | 10 | $^{-2}$ |
| **1,053.1** | ± | 4.0 | 8 | (± | 5 | )× | 10 | $^{-1}$ | **1,508.4** | ± | 0.4 | 7 | (± | 1 | )× | 10 | $^{0}$ |
| **1,066.1** | ± | 0.6 | 39 | (± | 4 | )× | 10 | $^{-1}$ | **1,513.3** | ± | 1.4 | 2 | (± | 2 | )× | 10 | $^{0}$ |
| **1,071.3** | ± | 2.0 | 1 | (± | 2 | )× | 10 | $^{-1}$ | **1,515.9** | ± | 0.5 | 4 | (± | 3 | )× | 10 | $^{-1}$ |
| **1,076.4** | ± | 1.5 | 39 | (± | 1 | )× | 10 | $^{-1}$ | **1,519.0** | ± | 1.5 | 20 | (± | 4 | )× | 10 | $^{-1}$ |
| **1,097.0** | ± | 3.1 | 10 | (± | 7 | )× | 10 | $^{-1}$ | **1,525.9** | ± | 2.4 | 8 | (± | 3 | )× | 10 | $^{-1}$ |
| **1,127.6** | ± | 2.2 | 5 | (± | 4 | )× | 10 | $^{-1}$ | **1,532.9** | ± | 1.6 | 2 | (± | 2 | )× | 10 | $^{-1}$ |
| **1,135.9** | ± | 0.3 | 15 | (± | 9 | )× | 10 | $^{-2}$ | **1,541.5** | ± | 1.5 | 26 | (± | 5 | )× | 10 | $^{-1}$ |
| **1,148.8** | ± | 4.4 | 3 | (± | 2 | )× | 10 | $^{-1}$ | **1,562.1** | ± | 1.2 | 58 | (± | 3 | )× | 10 | $^{-1}$ |
| **1,187.6** | ± | 0.1 | 29 | (± | 6 | )× | 10 | $^{-2}$ | **1,570.2** | ± | 1.0 | 10 | (± | 3 | )× | 10 | $^{0}$ |
| **1,192.1** | ± | 0.2 | 265 | (± | 4 | )× | 10 | $^{-2}$ | **1,579.6** | ± | 0.2 | 291 | (± | 3 | )× | 10 | $^{-2}$ |
| **1,201.4** | ± | 2.0 | 9 | (± | 4 | )× | 10 | $^{-1}$ | **1,595.7** | ± | 0.1 | 30 | (± | 3 | )× | 10 | $^{-1}$ |
| **1,211.1** | ± | 1.0 | 4 | (± | 5 | )× | 10 | $^{-1}$ | **1,610.4** | ± | 0.1 | 89 | (± | 3 | )× | 10 | $^{-2}$ |
| **1,217.8** | ± | 5.4 | 11 | (± | 8 | )× | 10 | $^{-1}$ | **1,633.5** | ± | 0.1 | 129 | (± | 2 | )× | 10 | $^{-2}$ |
| **1,231.8** | ± | 8.0 | 14 | (± | 8 | )× | 10 | $^{-1}$ | **1,647.1** | ± | 0.1 | 76 | (± | 2 | )× | 10 | $^{0}$ |
| **1,249.5** | ± | 1.7 | 19 | (± | 7 | )× | 10 | $^{-1}$ | **1,661.8** | ± | 0.2 | 54 | (± | 2 | )× | 10 | $^{-1}$ |
| **1,256.9** | ± | 0.3 | 1 | (± | 1 | )× | 10 | $^{-1}$ | **1,677.4** | ± | 0.1 | 5 | (± | 1 | )× | 10 | $^{-2}$ |
| **1,262.3** | ± | 1.0 | 3 | (± | 1 | )× | 10 | $^{-1}$ | **1,686.6** | ± | 0.1 | 668 | (± | 2 | )× | 10 | $^{-2}$ |
| **1,268.2** | ± | 1.4 | 14 | (± | 7 | )× | 10 | $^{-1}$ | **3,009.4** | ± | 4.8 | 6 | (± | 7 | )× | 10 | $^{0}$ |
| **1,288.9** | ± | 5.6 | 3 | (± | 3 | )× | 10 | $^{0}$ | **3,015.9** | ± | 2.1 | 12 | (± | 9 | )× | 10 | $^{0}$ |
| **1,299.7** | ± | 5.0 | 3 | (± | 3 | )× | 10 | $^{-1}$ | **3,022.9** | ± | 3.3 | 2 | (± | 1 | )× | 10 | $^{1}$ |
| **1,310.1** | ± | 3.3 | 18 | (± | 4 | )× | 10 | $^{-1}$ | **3,031.0** | ± | 5.6 | 2 | (± | 2 | )× | 10 | $^{1}$ |
| **1,316.5** | ± | 1.5 | 2 | (± | 1 | )× | 10 | $^{0}$ | **3,034.6** | ± | 2.6 | 1 | (± | 1 | )× | 10 | $^{1}$ |
| **1,331.6** | ± | 0.3 | 4 | (± | 2 | )× | 10 | $^{-1}$ | **3,037.6** | ± | 2.0 | 5 | (± | 4 | )× | 10 | $^{1}$ |
| **1,334.5** | ± | 0.6 | 12 | (± | 9 | )× | 10 | $^{-2}$ | **3,048.1** | ± | 6.7 | 1 | (± | 1 | )× | 10 | $^{1}$ |
| **1,342.0** | ± | 1.2 | 4 | (± | 2 | )× | 10 | $^{0}$ | **3,058.1** | ± | 6.6 | 7 | (± | 10 | )× | 10 | $^{0}$ |
| **1,345.4** | ± | 2.2 | 12 | (± | 8 | )× | 10 | $^{-1}$ | **3,068.9** | ± | 1.0 | 3 | (± | 2 | )× | 10 | $^{1}$ |
| **1,349.3** | ± | 2.3 | 3 | (± | 2 | )× | 10 | $^{0}$ | **3,070.8** | ± | 1.1 | 1 | (± | 2 | )× | 10 | $^{1}$ |
| **1,353.0** | ± | 1.0 | 7 | (± | 7 | )× | 10 | $^{0}$ | **3,081.3** | ± | 1.9 | 76 | (± | 9 | )× | 10 | $^{0}$ |
| **1,370.3** | ± | 4.2 | 7 | (± | 7 | )× | 10 | $^{0}$ | **3,099.9** | ± | 0.8 | 43 | (± | 3 | )× | 10 | $^{0}$ |
| **1,377.3** | ± | 0.4 | 138 | (± | 2 | )× | 10 | $^{-2}$ | **3,149.7** | ± | 0.4 | 60 | (± | 6 | )× | 10 | $^{-1}$ |
| **1,381.8** | ± | 3.2 | 9 | (± | 5 | )× | 10 | $^{0}$ | **3,154.3** | ± | 0.2 | 87 | (± | 9 | )× | 10 | $^{-1}$ |
| **1,393.2** | ± | 1.8 | 3 | (± | 2 | )× | 10 | $^{0}$ | **3,161.2** | ± | 0.1 | 118 | (± | 5 | )× | 10 | $^{-1}$ |
| **1,412.4** | ± | 2.1 | 2 | (± | 1 | )× | 10 | $^{0}$ | **3,173.8** | ± | 0.6 | 335 | (± | 4 | )× | 10 | $^{-2}$ |
| **1,417.2** | ± | 0.2 | 236 | (± | 4 | )× | 10 | $^{-2}$ | **3,178.4** | ± | 0.3 | 13 | (± | 2 | )× | 10 | $^{-2}$ |
| **1,424.3** | ± | 1.0 | 4 | (± | 4 | )× | 10 | $^{0}$ | **3,179.1** | ± | 0.1 | 43 | (± | 2 | )× | 10 | $^{0}$ |
| **1,430.1** | ± | 0.0 | 144 | (± | 4 | )× | 10 | $^{-2}$ | **3,179.8** | ± | 0.3 | 83 | (± | 4 | )× | 10 | $^{-1}$ |
| **1,439.8** | ± | 0.9 | 74 | (± | 2 | )× | 10 | $^{0}$ | **3,187.9** | ± | 0.1 | 34 | (± | 1 | )× | 10 | $^{0}$ |
| **1,450.8** | ± | 0.1 | 62 | (± | 2 | )× | 10 | $^{-1}$ | **3,188.8** | ± | 0.3 | 8 | (± | 4 | )× | 10 | $^{0}$ |
| **1,469.2** | ± | 0.2 | 10 | (± | 4 | )× | 10 | $^{0}$ | **3,189.3** | ± | 0.1 | 24 | (± | 4 | )× | 10 | $^{0}$ |
| **1,487.5** | ± | 2.4 | 16 | (± | 3 | )× | 10 | $^{0}$ | **3,205.3** | ± | 0.3 | 510 | (± | 6 | )× | 10 | $^{-1}$ |

| | | | | | | | |
|---|---|---|---|---|---|---|---|
| **3,205.6** | ± | 0.2 | 321 | (± | 5 | )× | 10 $^{-1}$ |
| **3,210.9** | ± | 6.2 | 173 | (± | 5 | )× | 10 $^{-1}$ |
| **3,223.0** | ± | 0.6 | 234 | (± | 1 | )× | 10 $^{-1}$ |
| **3,224.3** | ± | 0.2 | 234 | (± | 4 | )× | 10 $^{-1}$ |
| **3,239.0** | ± | 1.1 | 14 | (± | 1 | )× | 10 $^{0}$ |

a B3LYP/6-31G(d)

b Canonical ensemble average of five most stable configurations. The values following "±" sign is square root of Canonical ensemble variance.

c infrared

**Table S4**

Vibrational normal modes of AzCOCAn[a]

| Frequency [cm$^{-1}$] | | | IR Intensity [km/mol] | | | | | |
|---|---|---|---|---|---|---|---|---|
| 12.2 | ± | 1.1 | 61 | (± | 9 | )× | 10 | $^{-3}$ |
| 20.0 | ± | 0.9 | 9 | (± | 3 | )× | 10 | $^{-2}$ |
| 29.3 | ± | 1.8 | 11 | (± | 7 | )× | 10 | $^{-2}$ |
| 52.6 | ± | 3.2 | 2 | (± | 2 | )× | 10 | $^{-1}$ |
| 62.7 | ± | 2.8 | 8 | (± | 4 | )× | 10 | $^{-1}$ |
| 88.5 | ± | 3.0 | 10 | (± | 4 | )× | 10 | $^{-1}$ |
| 112.3 | ± | 2.5 | 72 | (± | 7 | )× | 10 | $^{-2}$ |
| 118.0 | ± | 0.6 | 5 | (± | 1 | )× | 10 | $^{-2}$ |
| 157.5 | ± | 8.8 | 35 | (± | 4 | )× | 10 | $^{-1}$ |
| 174.7 | ± | 0.8 | 16 | (± | 4 | )× | 10 | $^{-1}$ |
| 204.5 | ± | 7.6 | 15 | (± | 5 | )× | 10 | $^{-1}$ |
| 236.6 | ± | 1.0 | 4 | (± | 3 | )× | 10 | $^{-1}$ |
| 244.7 | ± | 4.3 | 4 | (± | 3 | )× | 10 | $^{0}$ |
| 258.3 | ± | 2.4 | 31 | (± | 6 | )× | 10 | $^{-1}$ |
| 295.4 | ± | 4.1 | 73 | (± | 5 | )× | 10 | $^{-1}$ |
| 327.6 | ± | 0.6 | 4 | (± | 2 | )× | 10 | $^{0}$ |
| 334.4 | ± | 3.3 | 32 | (± | 8 | )× | 10 | $^{-1}$ |
| 367.9 | ± | 3.1 | 4 | (± | 1 | )× | 10 | $^{0}$ |
| 391.4 | ± | 2.2 | 24 | (± | 3 | )× | 10 | $^{-1}$ |
| 397.6 | ± | 0.1 | 17 | (± | 6 | )× | 10 | $^{-1}$ |
| 412.0 | ± | 2.5 | 8 | (± | 4 | )× | 10 | $^{-1}$ |
| 415.7 | ± | 1.9 | 16 | (± | 5 | )× | 10 | $^{-1}$ |
| 424.3 | ± | 2.5 | 10 | (± | 6 | )× | 10 | $^{-1}$ |
| 431.5 | ± | 3.5 | 22 | (± | 7 | )× | 10 | $^{-1}$ |
| 451.8 | ± | 6.3 | 36 | (± | 8 | )× | 10 | $^{-1}$ |
| 486.3 | ± | 0.6 | 33 | (± | 7 | )× | 10 | $^{-2}$ |
| 508.0 | ± | 0.1 | 13 | (± | 8 | )× | 10 | $^{-2}$ |
| 527.3 | ± | 5.8 | 10 | (± | 2 | )× | 10 | $^{0}$ |
| 541.8 | ± | 5.7 | 10 | (± | 1 | )× | 10 | $^{0}$ |
| 558.5 | ± | 2.3 | 7 | (± | 3 | )× | 10 | $^{-1}$ |
| 586.0 | ± | 1.7 | 53 | (± | 4 | )× | 10 | $^{-1}$ |
| 591.2 | ± | 2.6 | 40 | (± | 4 | )× | 10 | $^{-1}$ |
| 615.5 | ± | 0.2 | 4 | (± | 2 | )× | 10 | $^{0}$ |
| 617.1 | ± | 0.6 | 6 | (± | 2 | )× | 10 | $^{0}$ |
| 653.6 | ± | 0.1 | 26 | (± | 3 | )× | 10 | $^{-1}$ |
| 660.5 | ± | 2.1 | 54 | (± | 2 | )× | 10 | $^{-1}$ |
| 670.5 | ± | 4.8 | 5 | (± | 1 | )× | 10 | $^{0}$ |
| 707.0 | ± | 2.7 | 6 | (± | 3 | )× | 10 | $^{0}$ |
| 722.4 | ± | 4.1 | 15 | (± | 4 | )× | 10 | $^{0}$ |
| 738.9 | ± | 0.5 | 1 | (± | 2 | )× | 10 | $^{0}$ |
| 749.4 | ± | 0.3 | 53 | (± | 4 | )× | 10 | $^{0}$ |
| 756.1 | ± | 0.1 | 9 | (± | 7 | )× | 10 | $^{-1}$ |
| 760.3 | ± | 1.5 | 24 | (± | 1 | )× | 10 | $^{0}$ |
| 772.8 | ± | 0.3 | 5 | (± | 3 | )× | 10 | $^{-1}$ |
| 788.8 | ± | 4.0 | 4 | (± | 1 | )× | 10 | $^{1}$ |
| 797.6 | ± | 1.1 | 14 | (± | 7 | )× | 10 | $^{0}$ |
| 801.7 | ± | 1.8 | 9 | (± | 3 | )× | 10 | $^{0}$ |
| 838.9 | ± | 0.4 | 6 | (± | 2 | )× | 10 | $^{-1}$ |
| 856.0 | ± | 0.8 | 8 | (± | 1 | )× | 10 | $^{0}$ |
| 867.8 | ± | 0.5 | 11 | (± | 7 | )× | 10 | $^{-1}$ |
| 877.9 | ± | 1.3 | 215 | (± | 9 | )× | 10 | $^{-2}$ |
| 880.7 | ± | 1.2 | 10 | (± | 2 | )× | 10 | $^{0}$ |
| 897.4 | ± | 2.1 | 4 | (± | 2 | )× | 10 | $^{0}$ |
| 905.1 | ± | 0.1 | 20 | (± | 3 | )× | 10 | $^{0}$ |
| 913.7 | ± | 2.7 | 9 | (± | 3 | )× | 10 | $^{0}$ |
| 919.9 | ± | 0.5 | 1 | (± | 2 | )× | 10 | $^{0}$ |
| 936.5 | ± | 7.7 | 8 | (± | 1 | )× | 10 | $^{0}$ |
| 955.9 | ± | 0.6 | 11 | (± | 3 | )× | 10 | $^{-1}$ |
| 962.6 | ± | 0.1 | 5 | (± | 2 | )× | 10 | $^{0}$ |
| 971.1 | ± | 0.8 | 4 | (± | 1 | )× | 10 | $^{0}$ |
| 985.0 | ± | 2.8 | 3 | (± | 3 | )× | 10 | $^{0}$ |
| 986.5 | ± | 2.3 | 21 | (± | 10 | )× | 10 | $^{-1}$ |
| 989.5 | ± | 1.1 | 25 | (± | 9 | )× | 10 | $^{-2}$ |
| 997.2 | ± | 4.0 | 5 | (± | 1 | )× | 10 | $^{-1}$ |
| 1,007.1 | ± | 5.2 | 3 | (± | 2 | )× | 10 | $^{0}$ |
| 1,015.2 | ± | 2.9 | 2 | (± | 1 | )× | 10 | $^{0}$ |
| 1,029.0 | ± | 2.1 | 8 | (± | 2 | )× | 10 | $^{0}$ |
| 1,037.3 | ± | 1.5 | 5 | (± | 3 | )× | 10 | $^{0}$ |
| 1,050.2 | ± | 0.2 | 57 | (± | 8 | )× | 10 | $^{-1}$ |
| 1,054.9 | ± | 2.8 | 3 | (± | 2 | )× | 10 | $^{1}$ |
| 1,066.7 | ± | 3.5 | 2 | (± | 1 | )× | 10 | $^{1}$ |
| 1,079.4 | ± | 2.1 | 18 | (± | 9 | )× | 10 | $^{0}$ |
| 1,085.4 | ± | 0.7 | 2 | (± | 2 | )× | 10 | $^{1}$ |
| 1,101.3 | ± | 1.6 | 258 | (± | 4 | )× | 10 | $^{0}$ |
| 1,136.3 | ± | 0.1 | 18 | (± | 5 | )× | 10 | $^{-2}$ |
| 1,185.6 | ± | 0.1 | 44 | (± | 7 | )× | 10 | $^{-2}$ |
| 1,192.0 | ± | 0.2 | 65 | (± | 3 | )× | 10 | $^{-1}$ |
| 1,205.1 | ± | 1.6 | 8 | (± | 2 | )× | 10 | $^{-1}$ |
| 1,213.5 | ± | 0.4 | 9 | (± | 2 | )× | 10 | $^{-1}$ |

| | | | | | | | |
|---|---|---|---|---|---|---|---|
| **1,220.5** | ± | 0.4 | 18 | (± | 6 | )× | 10 $^{-1}$ |
| **1,233.1** | ± | 1.7 | 7 | (± | 3 | )× | 10 $^{0}$ |
| **1,257.2** | ± | 0.5 | 6 | (± | 3 | )× | 10 $^{-1}$ |
| **1,265.6** | ± | 1.1 | 27 | (± | 1 | )× | 10 $^{-1}$ |
| **1,278.4** | ± | 5.8 | 4 | (± | 2 | )× | 10 $^{0}$ |
| **1,298.1** | ± | 0.3 | 8 | (± | 3 | )× | 10 $^{0}$ |
| **1,309.3** | ± | 3.4 | 2 | (± | 1 | )× | 10 $^{0}$ |
| **1,319.9** | ± | 2.6 | 3 | (± | 2 | )× | 10 $^{0}$ |
| **1,335.0** | ± | 0.8 | 13 | (± | 3 | )× | 10 $^{-1}$ |
| **1,345.1** | ± | 1.6 | 7 | (± | 6 | )× | 10 $^{0}$ |
| **1,358.1** | ± | 1.1 | 8 | (± | 3 | )× | 10 $^{0}$ |
| **1,374.6** | ± | 1.9 | 8 | (± | 2 | )× | 10 $^{-1}$ |
| **1,377.3** | ± | 0.8 | 13 | (± | 3 | )× | 10 $^{0}$ |
| **1,405.2** | ± | 3.3 | 7 | (± | 1 | )× | 10 $^{1}$ |
| **1,417.1** | ± | 0.6 | 21 | (± | 3 | )× | 10 $^{-2}$ |
| **1,425.8** | ± | 3.3 | 6 | (± | 3 | )× | 10 $^{0}$ |
| **1,432.4** | ± | 0.8 | 4 | (± | 2 | )× | 10 $^{0}$ |
| **1,443.2** | ± | 1.4 | 10 | (± | 1 | )× | 10 $^{1}$ |
| **1,452.1** | ± | 2.7 | 2 | (± | 1 | )× | 10 $^{0}$ |
| **1,465.7** | ± | 1.2 | 26 | (± | 3 | )× | 10 $^{0}$ |
| **1,488.0** | ± | 1.2 | 8 | (± | 7 | )× | 10 $^{0}$ |
| **1,496.8** | ± | 0.4 | 320 | (± | 9 | )× | 10 $^{-2}$ |
| **1,498.7** | ± | 0.2 | 39 | (± | 5 | )× | 10 $^{-1}$ |
| **1,509.0** | ± | 0.2 | 118 | (± | 8 | )× | 10 $^{-1}$ |
| **1,517.7** | ± | 1.7 | 3 | (± | 2 | )× | 10 $^{0}$ |
| **1,531.5** | ± | 2.9 | 27 | (± | 7 | )× | 10 $^{-1}$ |
| **1,555.7** | ± | 7.4 | 38 | (± | 3 | )× | 10 $^{-1}$ |
| **1,569.6** | ± | 2.1 | 56 | (± | 3 | )× | 10 $^{-1}$ |
| **1,581.7** | ± | 0.7 | 58 | (± | 4 | )× | 10 $^{-1}$ |
| **1,594.9** | ± | 0.3 | 34 | (± | 7 | )× | 10 $^{-1}$ |
| **1,610.6** | ± | 0.6 | 7 | (± | 2 | )× | 10 $^{-1}$ |
| **1,634.2** | ± | 0.1 | 22 | (± | 6 | )× | 10 $^{-2}$ |
| **1,646.8** | ± | 0.3 | 52 | (± | 2 | )× | 10 $^{0}$ |
| **1,659.7** | ± | 0.4 | 87 | (± | 7 | )× | 10 $^{-1}$ |
| **1,678.3** | ± | 0.3 | 104 | (± | 5 | )× | 10 $^{-2}$ |
| **1,687.4** | ± | 0.2 | 45 | (± | 2 | )× | 10 $^{-1}$ |
| **3,011.3** | ± | 12.7 | 47 | (± | 5 | )× | 10 $^{0}$ |
| **3,040.8** | ± | 24.9 | 4 | (± | 1 | )× | 10 $^{1}$ |
| **3,073.8** | ± | 1.5 | 23 | (± | 3 | )× | 10 $^{0}$ |
| **3,106.4** | ± | 8.0 | 18 | (± | 1 | )× | 10 $^{0}$ |
| **3,150.8** | ± | 0.3 | 766 | (± | 3 | )× | 10 $^{-2}$ |
| **3,158.2** | ± | 0.5 | 69 | (± | 6 | )× | 10 $^{-1}$ |
| **3,172.0** | ± | 1.3 | 5 | (± | 1 | )× | 10 $^{0}$ |
| **3,174.9** | ± | 1.2 | 16 | (± | 4 | )× | 10 $^{0}$ |
| **3,178.2** | ± | 0.1 | 18 | (± | 8 | )× | 10 $^{-2}$ |
| **3,179.6** | ± | 0.1 | 72 | (± | 3 | )× | 10 $^{-1}$ |
| **3,184.9** | ± | 1.0 | 38 | (± | 4 | )× | 10 $^{0}$ |
| **3,189.5** | ± | 0.1 | 15 | (± | 1 | )× | 10 $^{0}$ |
| **3,190.2** | ± | 0.1 | 26 | (± | 2 | )× | 10 $^{0}$ |
| **3,193.9** | ± | 0.9 | 15 | (± | 2 | )× | 10 $^{0}$ |
| **3,205.6** | ± | 0.1 | 55 | (± | 2 | )× | 10 $^{0}$ |
| **3,206.2** | ± | 0.3 | 26 | (± | 4 | )× | 10 $^{0}$ |
| **3,213.4** | ± | 4.5 | 14 | (± | 2 | )× | 10 $^{0}$ |
| **3,233.0** | ± | 2.3 | 144 | (± | 5 | )× | 10 $^{-1}$ |
| **3,236.0** | ± | 3.1 | 12 | (± | 2 | )× | 10 $^{0}$ |
| **3,240.5** | ± | 1.4 | 11 | (± | 3 | )× | 10 $^{0}$ |

a B3LYP/6-31G(d)

b Canonical ensemble average of five most stable configurations. The values following "±"sign is square root of Canonical ensemble variance.

c infrared

**Table S5**

Vibrational normal modes of AzCOCCOCAn[a]

| Frequency [cm⁻¹] | | | IR Intensity [km/mol] | | | | | |
|---|---|---|---|---|---|---|---|---|
| 7.0 | ± | 0.8 | 5 | (± | 2 | )× | 10 | -2 |
| 13.9 | ± | 1.2 | 5 | (± | 2 | )× | 10 | -2 |
| 18.4 | ± | 2.0 | 65 | (± | 10 | )× | 10 | -3 |
| 30.1 | ± | 0.5 | 19 | (± | 7 | )× | 10 | -2 |
| 37.5 | ± | 0.6 | 5 | (± | 1 | )× | 10 | -1 |
| 56.8 | ± | 1.9 | 5 | (± | 7 | )× | 10 | -1 |
| 72.2 | ± | 3.3 | 6 | (± | 2 | )× | 10 | -1 |
| 88.3 | ± | 3.5 | 14 | (± | 4 | )× | 10 | -1 |
| 112.3 | ± | 1.8 | 2 | (± | 1 | )× | 10 | -1 |
| 123.5 | ± | 2.7 | 5 | (± | 2 | )× | 10 | -1 |
| 134.7 | ± | 5.2 | 218 | (± | 9 | )× | 10 | -2 |
| 171.6 | ± | 3.0 | 13 | (± | 7 | )× | 10 | -1 |
| 182.9 | ± | 4.4 | 15 | (± | 4 | )× | 10 | -1 |
| 209.3 | ± | 8.5 | 5 | (± | 2 | )× | 10 | 0 |
| 230.1 | ± | 3.2 | 13 | (± | 5 | )× | 10 | -1 |
| 238.6 | ± | 1.4 | 3 | (± | 4 | )× | 10 | -1 |
| 258.4 | ± | 1.5 | 2 | (± | 2 | )× | 10 | 0 |
| 286.2 | ± | 5.1 | 9 | (± | 1 | )× | 10 | 0 |
| 297.3 | ± | 6.7 | 22 | (± | 2 | )× | 10 | -1 |
| 325.2 | ± | 2.4 | 27 | (± | 9 | )× | 10 | -1 |
| 332.3 | ± | 3.8 | 38 | (± | 7 | )× | 10 | -1 |
| 353.4 | ± | 2.1 | 3 | (± | 1 | )× | 10 | 0 |
| 382.5 | ± | 5.9 | 14 | (± | 4 | )× | 10 | -1 |
| 396.7 | ± | 0.8 | 14 | (± | 5 | )× | 10 | -1 |
| 399.5 | ± | 2.1 | 22 | (± | 9 | )× | 10 | -1 |
| 411.8 | ± | 2.0 | 18 | (± | 4 | )× | 10 | -1 |
| 419.1 | ± | 1.1 | 4 | (± | 2 | )× | 10 | 0 |
| 425.3 | ± | 3.5 | 2 | (± | 2 | )× | 10 | 0 |
| 440.5 | ± | 2.7 | 19 | (± | 7 | )× | 10 | -1 |
| 456.8 | ± | 3.7 | 4 | (± | 1 | )× | 10 | 0 |
| 484.7 | ± | 0.7 | 27 | (± | 7 | )× | 10 | -2 |
| 507.3 | ± | 1.1 | 1 | (± | 2 | )× | 10 | 0 |
| 519.8 | ± | 4.2 | 10 | (± | 4 | )× | 10 | 0 |
| 532.2 | ± | 1.8 | 14 | (± | 4 | )× | 10 | 0 |
| 551.3 | ± | 4.4 | 4 | (± | 4 | )× | 10 | 0 |
| 561.0 | ± | 3.3 | 4 | (± | 1 | )× | 10 | 0 |
| 587.2 | ± | 1.2 | 36 | (± | 2 | )× | 10 | -1 |
| 589.5 | ± | 0.5 | 55 | (± | 8 | )× | 10 | -1 |
| 615.8 | ± | 0.4 | 45 | (± | 10 | )× | 10 | -1 |
| 616.7 | ± | 0.3 | 3 | (± | 1 | )× | 10 | 0 |
| 654.4 | ± | 0.2 | 24 | (± | 3 | )× | 10 | -1 |
| 664.4 | ± | 3.6 | 87 | (± | 9 | )× | 10 | -1 |
| 675.2 | ± | 4.9 | 6 | (± | 2 | )× | 10 | 0 |
| 709.3 | ± | 2.5 | 8 | (± | 2 | )× | 10 | 0 |
| 719.1 | ± | 2.6 | 13 | (± | 2 | )× | 10 | 0 |
| 739.8 | ± | 0.6 | 2 | (± | 2 | )× | 10 | 0 |
| 749.0 | ± | 0.3 | 58 | (± | 2 | )× | 10 | 0 |
| 756.0 | ± | 0.4 | 20 | (± | 10 | )× | 10 | -1 |
| 761.6 | ± | 2.7 | 24 | (± | 1 | )× | 10 | 0 |
| 773.0 | ± | 0.3 | 11 | (± | 2 | )× | 10 | -1 |
| 787.0 | ± | 3.9 | 3 | (± | 1 | )× | 10 | 1 |
| 796.4 | ± | 0.5 | 11 | (± | 5 | )× | 10 | 0 |
| 799.3 | ± | 0.5 | 12 | (± | 3 | )× | 10 | 0 |
| 833.8 | ± | 8.1 | 3 | (± | 1 | )× | 10 | 0 |
| 845.6 | ± | 2.4 | 5 | (± | 1 | )× | 10 | 0 |
| 852.5 | ± | 2.9 | 8 | (± | 2 | )× | 10 | 0 |
| 861.1 | ± | 2.5 | 30 | (± | 10 | )× | 10 | -1 |
| 870.2 | ± | 1.7 | 32 | (± | 5 | )× | 10 | -1 |
| 879.3 | ± | 0.4 | 8 | (± | 2 | )× | 10 | 0 |
| 882.6 | ± | 4.3 | 7 | (± | 2 | )× | 10 | 0 |
| 898.4 | ± | 2.8 | 6 | (± | 4 | )× | 10 | 0 |
| 907.8 | ± | 2.6 | 20 | (± | 4 | )× | 10 | 0 |
| 920.6 | ± | 0.9 | 3 | (± | 1 | )× | 10 | 0 |
| 924.2 | ± | 1.9 | 5 | (± | 3 | )× | 10 | 0 |
| 953.3 | ± | 2.2 | 3 | (± | 5 | )× | 10 | 0 |
| 961.1 | ± | 2.6 | 4 | (± | 2 | )× | 10 | 0 |
| 968.6 | ± | 2.0 | 35 | (± | 3 | )× | 10 | -1 |
| 978.3 | ± | 2.6 | 37 | (± | 8 | )× | 10 | -1 |
| 984.9 | ± | 1.4 | 2 | (± | 1 | )× | 10 | 0 |
| 987.8 | ± | 0.5 | 5 | (± | 2 | )× | 10 | 0 |
| 991.6 | ± | 0.5 | 4 | (± | 2 | )× | 10 | 0 |
| 995.0 | ± | 0.9 | 6 | (± | 2 | )× | 10 | 0 |
| 1,001.9 | ± | 1.3 | 14 | (± | 7 | )× | 10 | -1 |
| 1,013.3 | ± | 1.5 | 10 | (± | 7 | )× | 10 | 0 |
| 1,019.7 | ± | 2.5 | 11 | (± | 4 | )× | 10 | 0 |
| 1,033.9 | ± | 1.4 | 7 | (± | 2 | )× | 10 | 0 |
| 1,038.0 | ± | 2.6 | 17 | (± | 4 | )× | 10 | 0 |
| 1,051.4 | ± | 0.3 | 3 | (± | 1 | )× | 10 | 0 |

| | | | | | | | |
|---|---|---|---|---|---|---|---|
| **1,055.7** | ± | 1.4 | 5 | (± | 1 | )× 10 | 0 |
| **1,067.8** | ± | 1.5 | 2 | (± | 1 | )× 10 | 1 |
| **1,072.9** | ± | 1.5 | 27 | (± | 7 | )× 10 | 0 |
| **1,080.0** | ± | 0.5 | 3 | (± | 3 | )× 10 | 0 |
| **1,122.9** | ± | 3.9 | 25 | (± | 5 | )× 10 | 1 |
| **1,135.6** | ± | 0.7 | 2 | (± | 2 | )× 10 | 1 |
| **1,147.1** | ± | 4.7 | 7 | (± | 3 | )× 10 | 1 |
| **1,170.0** | ± | 3.9 | 27 | (± | 7 | )× 10 | 0 |
| **1,186.9** | ± | 0.8 | 2 | (± | 2 | )× 10 | 0 |
| **1,192.1** | ± | 0.1 | 8 | (± | 1 | )× 10 | 0 |
| **1,205.2** | ± | 1.2 | 7 | (± | 2 | )× 10 | -1 |
| **1,214.1** | ± | 0.2 | 16 | (± | 2 | )× 10 | -1 |
| **1,220.4** | ± | 0.2 | 27 | (± | 3 | )× 10 | -1 |
| **1,234.8** | ± | 1.0 | 8 | (± | 3 | )× 10 | 0 |
| **1,257.6** | ± | 0.5 | 7 | (± | 4 | )× 10 | -1 |
| **1,266.7** | ± | 0.2 | 33 | (± | 4 | )× 10 | -1 |
| **1,282.7** | ± | 2.7 | 20 | (± | 3 | )× 10 | 0 |
| **1,293.3** | ± | 2.3 | 16 | (± | 8 | )× 10 | 0 |
| **1,300.4** | ± | 1.9 | 6 | (± | 4 | )× 10 | 0 |
| **1,310.7** | ± | 2.2 | 33 | (± | 5 | )× 10 | -1 |
| **1,319.9** | ± | 3.6 | 5 | (± | 5 | )× 10 | 0 |
| **1,334.8** | ± | 0.2 | 19 | (± | 6 | )× 10 | -1 |
| **1,338.9** | ± | 1.9 | 5 | (± | 1 | )× 10 | 0 |
| **1,346.8** | ± | 1.1 | 7 | (± | 5 | )× 10 | 0 |
| **1,361.7** | ± | 0.9 | 7 | (± | 2 | )× 10 | 0 |
| **1,371.2** | ± | 1.6 | 14 | (± | 7 | )× 10 | -1 |
| **1,376.5** | ± | 1.2 | 4 | (± | 2 | )× 10 | 0 |
| **1,399.9** | ± | 6.3 | 6 | (± | 2 | )× 10 | 1 |
| **1,415.6** | ± | 0.7 | 1 | (± | 1 | )× 10 | 1 |
| **1,416.8** | ± | 0.7 | 11 | (± | 5 | )× 10 | 0 |
| **1,429.9** | ± | 2.6 | 4 | (± | 6 | )× 10 | 0 |
| **1,435.0** | ± | 1.1 | 9 | (± | 4 | )× 10 | 0 |
| **1,441.8** | ± | 1.7 | 7 | (± | 2 | )× 10 | 1 |
| **1,447.6** | ± | 1.3 | 1 | (± | 2 | )× 10 | 1 |
| **1,450.4** | ± | 1.0 | 3 | (± | 2 | )× 10 | 0 |
| **1,467.0** | ± | 1.0 | 26 | (± | 2 | )× 10 | 0 |
| **1,488.6** | ± | 1.0 | 14 | (± | 5 | )× 10 | 0 |
| **1,496.0** | ± | 0.1 | 29 | (± | 2 | )× 10 | -1 |
| **1,498.4** | ± | 0.2 | 53 | (± | 5 | )× 10 | -1 |
| **1,508.6** | ± | 0.2 | 9 | (± | 2 | )× 10 | 0 |
| **1,510.6** | ± | 0.5 | 4 | (± | 1 | )× 10 | 0 |
| **1,521.4** | ± | 2.4 | 24 | (± | 7 | )× 10 | -1 |
| **1,526.8** | ± | 2.5 | 55 | (± | 5 | )× 10 | -1 |
| **1,541.2** | ± | 3.8 | 15 | (± | 9 | )× 10 | -1 |
| **1,547.7** | ± | 1.2 | 13 | (± | 4 | )× 10 | 0 |
| **1,568.2** | ± | 0.5 | 44 | (± | 5 | )× 10 | -1 |
| **1,580.5** | ± | 0.1 | 57 | (± | 2 | )× 10 | -1 |
| **1,595.2** | ± | 0.2 | 40 | (± | 5 | )× 10 | -1 |
| **1,609.8** | ± | 0.2 | 10 | (± | 1 | )× 10 | -1 |
| **1,634.0** | ± | 0.2 | 32 | (± | 2 | )× 10 | -2 |
| **1,647.3** | ± | 0.4 | 51 | (± | 2 | )× 10 | 0 |
| **1,660.4** | ± | 0.2 | 67 | (± | 7 | )× 10 | -1 |
| **1,677.7** | ± | 0.1 | 127 | (± | 8 | )× 10 | -2 |
| **1,687.1** | ± | 0.1 | 49 | (± | 1 | )× 10 | -1 |
| **2,980.0** | ± | 7.9 | 4 | (± | 1 | )× 10 | 1 |
| **3,015.9** | ± | 3.4 | 7 | (± | 1 | )× 10 | 1 |
| **3,035.2** | ± | 9.2 | 32 | (± | 9 | )× 10 | 0 |
| **3,051.1** | ± | 5.5 | 46 | (± | 5 | )× 10 | 0 |
| **3,063.2** | ± | 3.1 | 39 | (± | 8 | )× 10 | 0 |
| **3,079.3** | ± | 4.6 | 23 | (± | 8 | )× 10 | 0 |
| **3,099.2** | ± | 3.8 | 178 | (± | 8 | )× 10 | -1 |
| **3,115.1** | ± | 7.8 | 37 | (± | 4 | )× 10 | 0 |
| **3,150.1** | ± | 0.2 | 71 | (± | 4 | )× 10 | -1 |
| **3,157.4** | ± | 0.2 | 76 | (± | 10 | )× 10 | -1 |
| **3,171.7** | ± | 1.1 | 40 | (± | 3 | )× 10 | -1 |
| **3,173.9** | ± | 0.7 | 15 | (± | 4 | )× 10 | 0 |
| **3,178.7** | ± | 0.2 | 8 | (± | 3 | )× 10 | -1 |
| **3,180.2** | ± | 0.2 | 65 | (± | 3 | )× 10 | -1 |
| **3,183.7** | ± | 0.8 | 37 | (± | 2 | )× 10 | 0 |
| **3,189.7** | ± | 0.4 | 14 | (± | 3 | )× 10 | 0 |
| **3,190.9** | ± | 0.2 | 21 | (± | 4 | )× 10 | 0 |
| **3,192.9** | ± | 1.9 | 18 | (± | 3 | )× 10 | 0 |
| **3,206.3** | ± | 0.3 | 461 | (± | 10 | )× 10 | -1 |
| **3,207.0** | ± | 0.2 | 27 | (± | 2 | )× 10 | 0 |
| **3,212.4** | ± | 2.7 | 13 | (± | 3 | )× 10 | 0 |
| **3,227.3** | ± | 0.5 | 226 | (± | 7 | )× 10 | -1 |
| **3,239.1** | ± | 0.4 | 13 | (± | 1 | )× 10 | 0 |
| **3,243.6** | ± | 0.7 | 46 | (± | 4 | )× 10 | -1 |

a B3LYP/6-31G(d)

b Canonical ensemble average of five most stable configurations. The values following "±"sign is square root of Canonical ensemble variance.

c infrared

**Table S6**

Vibrational normal modes of CFC-113 calculated by an *ab initio* method (MP2/cc-pVTZ) using GAMESS.[a,b]

| MODE | 1 | 2 | 3 | 4 | 5 | 6 | 7 | 8 | 9 | 10 | 11 | 12 | 13 | 14 | 15 | 16 | 17 | 18 |
|---|---|---|---|---|---|---|---|---|---|---|---|---|---|---|---|---|---|---|
| Frequency[c] | 1212 | 1154 | 1074 | 1024 | 857 | 764 | 596 | 497 | 423 | 408 | 367 | 332 | 302 | 275 | 233 | 194 | 159 | 71 |

[a] Ref. 30.
[b] version March 24, 2007.
[c] The unit is $cm^{-1}$.